\documentclass[10pt,twocolumn,twoside]{IEEEtran}
\usepackage{setspace}
\usepackage{stfloats}
\usepackage{float}
\usepackage{cite}
\usepackage{psfig}
\usepackage{subfigure}
\usepackage[dvips]{graphicx}
\usepackage{amsmath}
\usepackage{amssymb}
\usepackage{multirow}
\newcommand{\bb}[1]{{\mathbf{#1}}}
\newcommand{\nn}{\nonumber}

\usepackage{mathtools}
\newlength{\arrow}
\newlength{\arrows}
\usepackage{color}

\begin{document}
\title{Fast Discrete Linear Canonical Transform Based on CM-CC-CM Decomposition and FFT}

\author{Soo-Chang Pei,~\IEEEmembership{Life Fellow,~IEEE,}
and Shih-Gu Huang
\thanks{
Copyright (c) 2015 IEEE. Personal use of this material is permitted. However, permission to use this material for any other purposes must be obtained from the IEEE by sending a request to pubs-permissions@ieee.org.

This work was supported by the Ministry of Science and Technology, Taiwan, under Contracts MOST 104-2221-E-002-096-MY3 and MOST 104-2221-E-002-006.

S.~C. Pei is with the Department of Electrical Engineering \& Graduate Institute of Communication Engineering, National Taiwan University, Taipei 10617, Taiwan (e-mail: peisc@ntu.edu.tw).

S.-G. Huang is with the Graduate Institute of Communication Engineering, National Taiwan University, Taipei 10617, Taiwan (e-mail: d98942023@ntu.edu.tw).
}
}

\maketitle
\begin{abstract}
In this paper, a discrete LCT (DLCT) irrelevant to the sampling periods and without oversampling operation is developed.
This DLCT is based on the well-known CM-CC-CM decomposition, that is, implemented by two discrete chirp multiplications (CMs) and one discrete chirp convolution (CC).
This decomposition doesn't use any scaling operation which will change the sampling period or cause
the interpolation error.
Compared with previous works, DLCT calculated by direct summation and DLCT based on center discrete dilated Hermite functions (CDDHFs), the proposed method implemented by FFTs has much lower computational complexity.
The relation between the proposed DLCT and the continuous LCT is also derived to approximate the samples of the continuous LCT.
Simulation results show that
the proposed method somewhat outperforms the CDDHFs-based method
in the approximation accuracy.
Besides, the proposed method has approximate additivity property with error as small as the CDDHFs-based method.
Most importantly, the proposed method has perfect reversibility, which doesn't hold in many existing DLCTs.
With this property, it is unnecessary to develop the inverse DLCT additionally because it can be replaced by the forward DLCT.
\end{abstract}

\begin{keywords}
ABCD transform, affine Fourier transform, fractional Fourier transform, linear canonical transform, quadratic-phase integrals.
\end{keywords}

\section{Introduction}\label{sec:Intro}
The linear canonical transform (LCT), first introduced in \cite{collins1970lens,moshinsky1971linear}, is a parameterized general linear integral transform with three degrees of freedom.
The LCT unifies a variety of transforms from  the well-known Fourier transform (FT),  fractional Fourier transform (FRFT) and Fresnel transform (also known as chirp convolution (CC)) to simple operations such as scaling and chirp multiplication (CM) \cite{ozaktas2001fractional,ding2001research,pei2002eigenfunctions}.
The LCT is an important tool in optics because 
the paraxial light propagation through a first-order optical system can be modeled by the LCT \cite{nazarathy1982first,bastiaans1989propagation,ozaktas2001fractional}.
Besides, as a generalization of the transforms mentioned above, the LCT could be more useful and attractive in many signal processing applications including filter design, radar  system  analysis, signal synthesis, time-frequency analysis, phase  reconstruction, pattern  recognition, graded index media analysis, encryption and modulation \cite{barshan1997optimal,pei2000simplified,pei2001relations,bastiaans2003phase,hennelly2005optical,sharma2006signal,pei2013reversible}.
In some papers, the LCT is also called Collins formula \cite{collins1970lens}, affine Fourier transform \cite{abe1994optical}, almost-Fresnel transformations \cite{abe1995almost}, generalized Fresnel transforms\cite{james1996generalized}, ABCD transforms \cite{bernardo1996abcd} or quadratic-phase integrals \cite{ozaktas2001fractional}.

In this paper, the definition of the LCT with four parameters \cite{ozaktas2001fractional,pei2011discrete,pei2013reversible} is used:
\begin{align}\label{eq:Intro04}
&X(u)\triangleq {\cal O}_{\textmd{LCT}}^{\bb M}\{x(t)\}\nn\\
&= \left\{
  \begin{array}{l l}
\sqrt {\frac{1}{{jb}}} \;\int\limits_{ - \infty }^\infty  {{e^{j2\pi \left( {\frac{d}{{2b}}{u^2} - \frac{1}{b}ut + \frac{a}{{2b}}{t^2}} \right)}}} x(t)\;dt,& \ b \ne 0\\
{\sqrt d \ {e^{j\pi \,cd\,{u ^2}}}x(du ),}& \ b = 0
  \end{array} \right.
\end{align}
where ${\cal O}_{\textmd{LCT}}^{\bb M}$ denotes the LCT operator and the $2\times2$ parameter matrix ${\bb M}$ is defined as
\begin{align}\label{eq:Intro08}
{\bb M} = (a,b;c,d) = \begin{bmatrix}
a\ &b\\
c\ &d
\end{bmatrix}\quad {\textmd{and}} \quad ad - bc = 1.
\end{align}
Although there are four parameters $(a,b;c,d)$, the degree of freedom is three according to
the constraint in (\ref{eq:Intro08}).
The LCT reduces to the FT when ${\bb M}=(0,1; -1,0)$, and becomes the FRFT when ${\bb M}=(\cos\alpha,\sin\alpha;-\sin\alpha,\cos\alpha)$. (Note that there is a constant phase difference between the LCT and FT/FRFT.
One can refer to \cite{pei2002eigenfunctions} for detailed description of the relations between the LCT and its special cases.)
The LCT has two important and useful properties:
reversibility and additivity  \cite{pei2002eigenfunctions,pei2011discrete}.
The reversibility property allows one to realize the inverse LCT (ILCT) with parameter matrix ${\bb M}$ by the forward LCT with parameter matrix ${\bb M}^{-1}$:
\begin{align}\label{eq:Intro12}
{\cal O}_{\textmd{ILCT}}^{\bb M} \buildrel \Delta \over = {\left[ {{\cal O}_{\textmd{LCT}}^{\bb M}} \right]^{ - 1}} = {\cal O}_{\textmd{LCT}}^{{{\bb M}^{ - 1}}}.
\end{align}
The additivity property of the LCT is given by
\begin{align}\label{eq:Intro16}
{\cal O}_{\textmd{LCT}}^{{{\bb M}_1}}\begin{array}{*{20}{l}}
{\cal O}_{\textmd{LCT}}^{{{\bb M}_2}} = {\cal O}_{\textmd{LCT}}^{{{\bb M}_1} \times {{\bb M}_2}}.
\end{array}
\end{align}
It implies that the cascade of several LCTs with parameter matrices ${\bb M}_1,{\bb M}_2,\ldots,{\bb M}_k$ can be replaced by only one LCT using parameter matrix ${\bb M}={\bb M}_1\times {\bb M}_2\times\cdots\times {\bb M}_k$.
In other words, the LCT can be decomposed into a cascade of multiple LCTs.
The
reversibility property is a special case of the additivity property when ${\bb M}_1={\bb M}_2^{-1}$.
Besides, it is worth noting that the LCT is not commutative
in most cases, i.e. ${\cal O}_{\textmd{LCT}}^{{{\bb M}_1}}{\cal O}_{\textmd{LCT}}^{{{\bb M}_2}} \ne {\cal O}_{\textmd{LCT}}^{{{\bb M}_2}}{\cal O}_{\textmd{LCT}}^{{{\bb M}_1}}$, because ${{\bb M}_1} {{\bb M}_2} \ne {{\bb M}_2} {{\bb M}_1}$.

Consider a set of sampled data $x[n]$ with sampling period $\Delta_t$, i.e. $x[n]=x(n\Delta_t)$. As $b\neq0$, the sampled output of the LCT $X[k]=X(k\Delta_u)$ is given by
\begin{align}\label{eq:Intro20}
X[k]= \sqrt {\frac{1}{{jb}}} \sum\limits_n^{} {{e^{j2\pi \left( {\frac{d}{{2b}}{k^2}\Delta _u^2 - \frac{1}{b}kn{\Delta _u}{\Delta _t} + \frac{a}{{2b}}{n^2}\Delta _t^2} \right)}}} x[n]{\Delta _t}.
\end{align}
This direct
summation method is inefficient because of its high computational complexity.
In \cite{zhao2013unitary}, the authors proposed some conditions for the sampling periods such that  (\ref{eq:Intro20}) becomes an unitary discrete transform, however, no low-complexity implementation method is proposed.
Taking the benefit of the additivity property mentioned in (\ref{eq:Intro16}),
the complexity can be reduced by decomposing the LCT into a sequence of
simpler operations, such as scaling, chirp multiplication (CM), chirp convolution (CC) and FRFT \cite{papoulis1977signal,hennelly2005fast,ozaktas2006efficient,koc2008digital,pei2011discrete}.
In\cite{papoulis1977signal}, two kinds of decompositions are proposed, decomposing the LCT into CMs and CCs.
But there is no further discussion on the digital implementation based on the decompositions.
Most of the other kinds of decompositions are introduced in \cite{koc2008digital}.
The first method in \cite{koc2008digital} decomposes
the LCT into scaling, CMs and Fourier transforms.
The second 
method is based on the well-known Iwasawa decomposition \cite{wolf2004geometric,simon1998iwasawa}, i.e. one FRFT, one scaling and one CM.
It follows that the digital computation of the LCT can be realized by discrete CM, discrete Fourier transform (DFT) and/or discrete FRFT (DFRFT).
It has been shown in \cite{pei2001relations,koc2008digital} that the Wigner distribution function (WDF) of the LCT output is a linearly affine distorted version of the input.
The bandwidth-time
(BT) product of the output would be larger than the input.
To avoid aliasing effect, oversampling is utilized to increase the number of samples (i.e. reduce the sampling period).
The two methods in \cite{koc2008digital} require smaller oversampling rate than the direct computation in (\ref{eq:Intro20}), and thus can further lower the complexity.
In \cite{pei2011discrete}, a DLCT with no oversampling involved is proposed, which is also based on the Iwasawa decomposition.
Since the Hermite functions are the eigenfunctions of the FRFT \cite{pei2002eigenfunctions}, the DFRFT can be implemented by an orthonormal basis of discrete Hermite functions.
Since Iwasawa decomposition involves a scaling operation, the scalable discrete Hermite functions, called center discrete dilated Hermite functions (CDDHFs) \cite{pei2012signal}, are used to implement the DLCT.
The oversampling operation is
not used in this DLCT.
So the length of output signal will remain the same as the input.
Besides, simulation results in \cite{pei2011discrete} show that this DLCT has smaller error in the reversibility and additivity properties in comparison with Ko{\c{c}}'s methods \cite{koc2008digital}.
However, the main disadvantage is the high computational complexity on the generation of CDDHFs.

Actually, Ko{\c{c}}'s work \cite{koc2008digital} is
mainly focused on determining the sampling periods and number of samples so that the continuous LCT can be recovered from its samples.
In this paper, the focus is on the development of a discrete LCT (DLCT) which is irrelevant to the sampling periods and doesn't
involve oversampling.
Considering the FT, i.e. a special case of the LCT, it is expected that the DLCT can reduce to the DFT in this special case.
Sampling periods and aliasing effect are not the concerns of the DFT, and thus also not of the DLCT.
Since scaling operation will change the sampling period \cite{koc2008digital} or cause
the interpolation error,
the CM-CC-CM decomposition
\cite{papoulis1977signal,ozaktas2001fractional} that involves no scaling is adopted in this paper.
The CM-CC-CM decomposition is also used in the implementations of many other transforms such as FRFT \cite{ozaktas1996digital,ozaktas2001fractional}, complex LCT \cite{kocc2011fast,ozaktas2015fast} and gyrator transform \cite{pei2015discrete}.
The proposed DLCT based on CM-CC-CM decomposition consists of three discrete CMs, one DFT and one IDFT.
In comparison with the DLCT based on CDDHFs \cite{pei2011discrete}, which is also irrelevant to sampling periods, the proposed DLCT has much less computational complexity.
Besides, simulation results show that in some cases, the proposed DLCT can approximate the samples of the continuous LCT with higher accuracy and yield smaller error in the additivity property.
Most importantly, the proposed DLCT has ``perfect'' reversibility and thus is more useful in some applications such as encryption/decryption.

This paper is organized as follows.
Section~\ref{sec:Lai} gives a review of the DLCT proposed by Pei and Lai \cite{pei2011discrete} for comparison with the newly proposed DLCT.
In Section~\ref{sec:CMCCCM}, we develop a DLCT based on the CM-CC-CM decomposition.
Some special cases of the proposed DLCT are discussed.
We also investigate the relation between the proposed DLCT and the continuous LCT
in Section~\ref{sec:Relation}.
In Section~\ref{sec:Comp}, the proposed DLCT will be compared with
the previous work \cite{pei2011discrete}.
Finally, conclusions are 
made in Section~\ref{sec:Con}.

\section{Review of DLCT Based on Center Discrete Dilated Hermite Functions \cite{pei2011discrete}}\label{sec:Lai}
The goal of this paper is on the development of a discrete LCT (DLCT) which is irrelevant to the sampling periods and without oversampling procedure.
We first give a brief review of the DLCT in \cite{pei2011discrete}, which will be used for comparison with the newly proposed DLCT.
In \cite{pei2011discrete}, the 
DLCT is based on the Iwasawa decomposition \cite{wolf2004geometric,simon1998iwasawa}:
\begin{align}\label{eq:Lai04}
{\bb M}=\begin{bmatrix}
a&b\\
c&d
\end{bmatrix}
=\begin{bmatrix}
1&0\\
\xi&1
\end{bmatrix}
\begin{bmatrix}
\sigma&0\\
0&\sigma^{-1}
\end{bmatrix}
\begin{bmatrix}
\cos\alpha&\sin\alpha\\
-\sin\alpha&\cos\alpha
\end{bmatrix}.
\end{align}
The first matrix corresponds to chirp multiplication (CM) with chirp rate $\xi=(ac+bd)/(a^2+b^2)$.
The second matrix corresponds to scaling operation with scaling parameter $\sigma=\sqrt{a^2+b^2}$.
The last one corresponds to FRFT with fractional angle $\alpha$ that satisfies $\cos\alpha=a/\sigma$ and $\sin\alpha=b/\sigma$.

It has been known that Hermite functions (HFs) are the eigenfunctions of the FRFT \cite{papoulis1962fourier,namias1980fractional}.
From (\ref{eq:Lai04}), if the input is first expanded by the HFs, the LCT output can be expressed by the scaled HFs multiplied by a chirp function.
In order to develop a unitary DLCT, orthonormal discrete HFs (DHFs) and orthonormal scalable DHFs are necessary.
Therefore, Pei and Lai utilize the center discrete dilated Hermite functions (CDDHFs) \cite{pei2012signal}, which are orthonormal and can approximate the samples of the scaled HFs.
Let $\Phi_{p;\sigma}^{\textmd{CDDHF}}[n]$ denote the CDDHF of order $p$ with scaling parameter $\sigma$.
Then, $\Phi_{p;1}^{\textmd{CDDHF}}[n]$ is the undilated DHF of order $p$.
The DFRFT and discrete scaling are designed satisfying
\begin{align}
{\cal O}_{\textmd{DLCT}}^{(\cos\alpha ,\sin\alpha ; - \sin\alpha ,\cos\alpha )}&\left\{ {\Phi _{p;1}^{{\textmd{CDDHF}}}[n]} \right\} = {e^{ - j\left( {p + \frac{1}{2}} \right)\alpha }} \nn\\
&\qquad\qquad\qquad\qquad\cdot \Phi _{p;1}^{{\textmd{CDDHF}}}[n],\label{eq:Lai08}\\
{\cal O}_{{\textmd{DLCT}}}^{(\sigma ,0;{\textmd{ }}0,{\sigma ^{ - 1}})}&\left\{ {\Phi _{p;1}^{{\textmd{CDDHF}}}[n]} \right\} = \Phi _{p;\sigma }^{{\textmd{CDDHF}}}[n]\label{eq:Lai12},
\end{align}
respectively.
Denote ${\bf{E}}_\sigma$ as an $N\times N$ orthonormal matrix with $p$th column being $\Phi_{p;\sigma}^{\textmd{CDDHF}}[n]$.
From (\ref{eq:Lai08}) and (\ref{eq:Lai12}), the matrix-computational expressions of the discrete CM, discrete scaling and DFRFT in (\ref{eq:Lai04}) are ${{\bf{C}}_\xi }$, $ {{{\bf{E}}_\sigma }{\bf{E}}_1^T}$ and ${{{\bf{E}}_1}{{\bf{V}}_\alpha }{\bf{E}}_1^T}$, respectively,
where ${\bf{V}}_\alpha$ and ${\bf{C}}_\xi$ are $N\times N$ diagonal matrices consisting of
 the eigenvalues in (\ref{eq:Lai08}) and chirp samples, respectively:
\begin{align}\label{eq:Lai20}
{\left[ {{{\bf{V}}_\alpha }} \right]_{p,p}} = {e^{ - j\left( {p + \frac{1}{2}} \right)\alpha }},
\quad{\left[ {{{\bf{C}}_\xi }} \right]_{k,k}} =
{e^{j\frac{\pi }{N}\xi k^2}}.
\end{align}
Therefore, the CDDHFs-based DLCT is given by
\begin{align}\label{eq:Lai16}
{\bf{X}} = {\cal O}_{{\textmd{DLCT}}}^{\bb M}\left\{ {\bf{x}} \right\} &= {{\bf{C}}_\xi }\left( {{{\bf{E}}_\sigma }{\bf{E}}_1^T} \right)\left( {{{\bf{E}}_1}{{\bf{V}}_\alpha }{\bf{E}}_1^T} \right){\bf{x}}\nn\\
&= {{\bf{C}}_\xi }{{\bf{E}}_\sigma }{{\bf{V}}_\alpha }{\bf{E}}_1^T{\bf{x}},
\end{align}
where $\bb x$ and $\bb X$ are $N\times1$ vectors consisting of $x[n]$ and the DLCT output $X[k]$, respectively.

This DLCT has three main disadvantages.
First, simulation results show that this DLCT can approximate the continuous LCT well; however, the sampling periods are restricted to $\Delta_t=\Delta_u=\sqrt{1/N}$ that is used when generating the CDDHFs.
Second, it doesn't satisfy the reversibility property perfectly, even though the error is smaller than other previous works.
The last one 
is the relatively high computational complexity, even higher than the direct computation in (\ref{eq:Intro20}). 
Besides, the generation of CDDHFs (i.e. ${\bf{E}}_\sigma$) is based on eigendecomposition, which is time-consuming, and precomputing ${\bf{E}}_\sigma$ is impractical because different ${\bf{E}}_\sigma$ is used for different $\sigma$ (different parameter matrix $\bb M$).

\section{DLCT Based on CM-CC-CM Decomposition}\label{sec:CMCCCM}
In this paper, the CM-CC-CM decomposition
\cite{papoulis1977signal,ozaktas2001fractional} is utilized.
The digital computation of the LCT based on this decomposition has been mentioned by Ko{\c{c}} \emph{et al.} in one brief paragraph of \cite{koc2008digital}.
However, their work concentrates on the oversampling operation in each step so that the number of samples is
sufficient for recovering the corresponding continuous signal.
The development of DLCT, which is irrelevant to the sampling periods and doesn't involve oversampling, for discrete data is the main object of our interest.
In short, Ko{\c{c}}'s work is like digitally computing the FT with careful attention to the sampling periods, while our work is much like the development of DFT.

Although a variety of decompositions have been presented in \cite{koc2008digital}, we adopt the CM-CC-CM decomposition because it doesn't have scaling operation.
In \cite{koc2008digital}, scaling merely changes the sampling period and reinterprets the same samples with the scaled period; that is, 
the output of $x[n]=x(n\Delta)$ is $X[k]=x[k]=x(kd\Delta)$.

\subsection{Formulation of the DLCT For $B\neq0$}\label{subsec:Bnot0}
We directly develop the DLCT in the discrete domain.
Firstly, 
consider the DLCT 
based on direct summation:
\begin{align}
\bb X&={\cal O}_{\textmd{DLCT}}^{\bb M}\{\bb x\} = \bb K\bb x,\label{eq:Bnot002}\\
\textmd{i.e.}\ \ X[k]&= \sqrt {\frac{1}{{jBN}}} \sum\limits_{n =  - N/2 }^{N/2-1}  {{e^{j\frac{{2\pi }}{N}\left( {\frac{D}{{2B}}{k^2} - \frac{1}{B}kn + \frac{A}{{2B}}{n^2}} \right)}}} x[n],\label{eq:Bnot004}
\end{align}
where $\bb x$ and $\bb X$ are $N\times1$ vectors consisting of $x[n]$ and the DLCT output $X[k]$, respectively; $\bb K$ is the DLCT kernel of size $N\times N$; and $\bb M$ is the $2\times2$ parameter matrix defined as
${\bb M} = (A,B;C,D)$.
The range of $n$ in the summation of (\ref{eq:Bnot004}), i.e. $[-N/2,N/2-1]$, is replaced by $[-(N-1)/2,(N-1)/2]$ if $n$ is odd.
Note that $(A,B;C,D)$ would be different from the parameters $(a,b;c,d)$ used in continuous LCT.
Later we will show that $(A,B;C,D)$ depends on $(a,b;c,d)$ and the sampling period $\Delta$.
It is obvious that (\ref{eq:Bnot004}) reduces to the DFT multiplied by a constant phase $\sqrt{-j}$ when $M=(0,1;-1,0)$.
If we want the inverse DLCT (IDLCT) 
given by
\begin{align}
\bb x&={\cal O}_{\textmd{DLCT}}^{\bb M^{-1}}\{\bb X\} = \bb K^\dag\bb x,\label{eq:Bnot010}\\
\textmd{i.e.}\ \ x[n]&= \sqrt {\frac{1}{{-jBN}}} \sum\limits_{k =  - N/2 }^{N/2-1}  {{e^{j\frac{{2\pi }}{N}\left( { -\frac{A}{{2B}}{n^2}}+ \frac{1}{B}kn - \frac{D}{{2B}}{k^2} \right)}}} X[k],\label{eq:Bnot012}
\end{align}
it is required that  ${\bb M}^{-1} = (D,-B;-C,A)$ and
\begin{align}\label{eq:Bnot016}
AD-BC=1.
\end{align}
Unfortunately, in other cases, this definition usually violates the reversibility property, i.e. $x[n]\neq{\cal O}_{\textmd{DLCT}}^{\bb M^{-1}}\left\{{\cal O}_{\textmd{DLCT}}^{\bb M}\{x[n]\}\right\}$.
Besides, the direct summation is quite inefficient because it involves $N^2$ complex multiplications.

Accordingly, we introduce the CM-CC-CM decomposition:
\begin{align}\label{eq:Bnot020}
{\bb M}=\begin{bmatrix}
A&B\\
C&D
\end{bmatrix}
=
\begin{bmatrix}
1&0\\
\frac{D-1}{B}&1
\end{bmatrix}
\begin{bmatrix}
1&B\\
0&1
\end{bmatrix}
\begin{bmatrix}
1&0\\
\frac{A-1}{B}&1
\end{bmatrix}.
\end{align}
These three matrices from left to right represent chirp multiplication (CM) with chirp rate $(D-1)/B$, chirp convolution (CC) with parameter $B$, and again CM with chirp rate $(A-1)/B$, respectively.
The CC can be further decomposed into IDFT-CM-DFT, i.e.
\begin{align}\label{eq:Bnot024}
\begin{bmatrix}
1&B\\
0&1
\end{bmatrix}
=
\begin{bmatrix}
0&-1\\
1&0
\end{bmatrix}
\begin{bmatrix}
1&0\\
-B&1
\end{bmatrix}
\begin{bmatrix}
0&1\\
-1&0
\end{bmatrix}.
\end{align}
Then, we can develop a DLCT completely composed of DFT, IDFT and discrete CMs.
The three steps of computation of the proposed DLCT are shown below:
\begin{align}
&\textmd{(CM)}\ x_1[n]={e^{j\frac{\pi }{N}\frac{{A - 1}}{B}{n^2}}}x[n] \label{eq:Bnot028},\\
&\textmd{(CC)}\ X_1[k]=\frac{1}{N}\sum\limits_{m =  - N/2 }^{N/2-1}  {e^{ - j\frac{\pi }{N}B{m^2}}}\nn\\
&\qquad\qquad\qquad\quad\cdot\left(\sum\limits_{n =  - N/2 }^{N/2-1}  {{x_1}[n]{e^{ - j\frac{{2\pi }}{N}mn}}} \right) {e^{  j\frac{{2\pi }}{N}mk}} \label{eq:Bnot032},\\
&\textmd{(CM)}\ X[k]={e^{j\frac{\pi }{N}\frac{{D - 1}}{B}{k^2}}}X_1[k] \label{eq:Bnot036}.
\end{align}
Denote $\bb F$ and $\bb F^\dag$ as the DFT and IDFT matrices, and $\bb C_\xi$ as a diagonal matrix where the $(n,n)$-th element is ${\left[ {{{\bf{C}}_\xi }} \right]_{n,n}} ={e^{j\frac{\pi }{N}\xi n^2}}$.
Then, the matrix form of the DLCT is given by
\begin{align}\label{eq:Bnot040}
{\bf{X}} = {\cal O}_{{\textmd{DLCT}}}^{\bb M}\left\{ {\bf{x}} \right\} = {{\bf{C}}_{\frac{{D - 1}}{B}}}{{\bf{F}}^\dag }{{\bf{C}}_{ - B}}{\bf{F}}{{\bf{C}}_{\frac{{A - 1}}{B}}}{\bf{x}}. 
\end{align}
Since the DFT/IDFT can be implemented by fast Fourier transform (FFT), the computational complexity is apparently lower than  that of the direct summation method (\ref{eq:Bnot004}).
Replacing $\bb M$ in (\ref{eq:Bnot040}) by $\bb M^{-1}$ leads to the IDLCT based on CM-CC-CM decomposition:
\begin{align}\label{eq:Bnot044}
{\bf{x}} ={\cal O}_{{\textmd{DLCT}}}^{\bb M^{-1}}\left\{ {\bf{X}} \right\} = {{\bf{C}}_{ - \frac{{A - 1}}{B}}}{{\bf{F}}^\dag }{{\bf{C}}_B}{\bf{F}}{{\bf{C}}_{ - \frac{{D - 1}}{B}}}{\bf{X}}.
\end{align}
It is obvious that the proposed DLCT satisfies the reversibility property because
\begin{align}\label{eq:Bnot048}
\bb x&={{\bf{C}}_{ - \frac{{A - 1}}{B}}}{{\bf{F}}^\dag }{{\bf{C}}_B}{\bf{F}}{{\bf{C}}_{ - \frac{{D - 1}}{B}}}
\left({{\bf{C}}_{\frac{{D - 1}}{B}}}{{\bf{F}}^\dag }{{\bf{C}}_{ - B}}{\bf{F}}{{\bf{C}}_{\frac{{A - 1}}{B}}}{\bf{x}}\right)\nn\\
&={\cal O}_{{\textmd{DLCT}}}^{\bb M^{-1}}\left\{ {{\cal O}_{{\textmd{DLCT}}}^{\bb M}\left\{ {\bf{x}} \right\}} \right\}.
\end{align}

\subsection{Formulation of the DLCT For $B=0$}\label{subsec:Bis0}
The definition in (\ref{eq:Bnot040}) is invalid when $B=0$.
Fortunately, if $B=0$, one has
$A\neq0$ and $D\neq0$ because $AD-BC=AD=1$.
In this case, the following two kinds of decompositions are considered:
\begin{align}\label{eq:Bis004}
\bb M
= \begin{bmatrix}
{{A}}&0\\
C&D
\end{bmatrix}
&=\begin{bmatrix}
0&1\\
{ - 1}&0
\end{bmatrix}
\begin{bmatrix}
{ - C}&{ - D}\\
{{A}}&0
\end{bmatrix}\nn\\
&=\begin{bmatrix}
0&1\\
{ - 1}&0
\end{bmatrix}
\begin{bmatrix}
1&0\\
{\frac{1}{D}}&1
\end{bmatrix}
\begin{bmatrix}
1&{ - D}\\
0&1
\end{bmatrix}
\begin{bmatrix}
1&0\\
{\frac{{C + 1}}{D}}&1
\end{bmatrix},
\end{align}
and
\begin{align}\label{eq:Bis008}
\bb M
= \begin{bmatrix}
{{A}}&0\\
C&D
\end{bmatrix}
 &= \begin{bmatrix}
0&A\\
{ - D}&C
\end{bmatrix}
\begin{bmatrix}
0&{ - 1}\\
1&0
\end{bmatrix}\nn\\
 &= \begin{bmatrix}
1&0\\
{\frac{{C - 1}}{A}}&1
\end{bmatrix}
\begin{bmatrix}
1&A\\
0&1
\end{bmatrix}
\begin{bmatrix}
1&0\\
{ - \frac{1}{A}}&1
\end{bmatrix}
\begin{bmatrix}
0&{ - 1}\\
1&0
\end{bmatrix}.
\end{align}
The matrix forms of the DLCTs based on (\ref{eq:Bis004}) and (\ref{eq:Bis008}) are given by
\begin{align}\label{eq:Bis012}
{\bf{X}} &={\cal O}_{{\textmd{DLCT}}}^{(A,0;C,D)}\left\{ {\bf{x}} \right\} = \sqrt{-j}\ {\bf{F}} {{\bf{C}}_{\frac{1}{D}}}{{\bf{F}}^\dag }{{\bf{C}}_D}{\bf{F}}{{\bf{C}}_{\frac{{C + 1}}{D}}}{\bf{x}},\\
{\bf{X}} &= {\cal O}_{{\textmd{DLCT}}}^{(A,0;C,D)}\left\{ {\bf{x}} \right\} = \sqrt{j}\ {{\bf{C}}_{\frac{{C - 1}}{A}}}{{\bf{F}}^\dag }{{\bf{C}}_{ - A}}{\bf{F}}{{\bf{C}}_{ - \frac{1}{A}}}{{\bf{F}}^\dag }{\bf{x}}\label{eq:Bis016},
\end{align}
respectively.
Note that $\sqrt{-j}$ (or $\sqrt{j}$)  is the phase difference between LCT and FT (or inverse FT).

The reversibility property for $B=0$ is given by
\begin{align}\label{eq:Bis017}
{\bf{x}} ={\cal O}_{{\textmd{DLCT}}}^{(D,0;-C,A)}\left\{ {{\cal O}_{{\textmd{DLCT}}}^{(A,0;C,D)}\left\{ {\bf{x}} \right\}} \right\}.
\end{align}
If both the forward and inverse transforms use the same DLCT form in (\ref{eq:Bis012}) or (\ref{eq:Bis016}), the reversibility property doesn't holds
due to no intermediate cancellation in between:
\begin{align}\label{eq:Bis018_1}
{\bf{x}} &\neq \sqrt{-j}\ {\bf{F}} {{\bf{C}}_{\frac{1}{A}}}{{\bf{F}}^\dag }{{\bf{C}}_A}{\bf{F}}{{\bf{C}}_{\frac{{-C + 1}}{A}}}\nn\\
  &\qquad\qquad\qquad\qquad\left\{\sqrt{-j}\ {\bf{F}} {{\bf{C}}_{\frac{1}{D}}}{{\bf{F}}^\dag }{{\bf{C}}_D}{\bf{F}}{{\bf{C}}_{\frac{{C + 1}}{D}}}{\bf{x}}\right\}\\
 &\neq\sqrt{j}\ {{\bf{C}}_{\frac{{-C - 1}}{D}}}{{\bf{F}}^\dag }{{\bf{C}}_{ - D}}{\bf{F}}{{\bf{C}}_{ - \frac{1}{D}}}{{\bf{F}}^\dag }\nn\\
 &\qquad\qquad\qquad\quad\left\{\sqrt{j}\ {{\bf{C}}_{\frac{{C - 1}}{A}}}{{\bf{F}}^\dag }{{\bf{C}}_{ - A}}{\bf{F}}{{\bf{C}}_{ - \frac{1}{A}}}{{\bf{F}}^\dag }{\bf{x}}\right\}\label{eq:Bis018_2}.
\end{align}
However, if we let the forward and inverse transforms use different DLCT forms, reversibility can be achieved
because ${\bf{F}}^\dag \bb F=\bb F {\bf{F}}^\dag=\bb I$ and $\bb C_\xi\bb C_{-\xi}=\bb I$ in between:
\begin{align}\label{eq:Bis019_1}
{\bf{x}} &= \sqrt{j}\ {{\bf{C}}_{\frac{{-C - 1}}{D}}}{{\bf{F}}^\dag }{{\bf{C}}_{ - D}}{\bf{F}}{{\bf{C}}_{ - \frac{1}{D}}}{{\bf{F}}^\dag }\nn\\
  &\qquad\qquad\qquad\qquad\left\{\sqrt{-j}\ {\bf{F}} {{\bf{C}}_{\frac{1}{D}}}{{\bf{F}}^\dag }{{\bf{C}}_D}{\bf{F}}{{\bf{C}}_{\frac{{C + 1}}{D}}}{\bf{x}}\right\}\\
 &=\sqrt{-j}\ {\bf{F}} {{\bf{C}}_{\frac{1}{A}}}{{\bf{F}}^\dag }{{\bf{C}}_A}{\bf{F}}{{\bf{C}}_{\frac{{-C + 1}}{A}}}\nn\\
 &\qquad\qquad\qquad\quad\left\{\sqrt{j}\ {{\bf{C}}_{\frac{{C - 1}}{A}}}{{\bf{F}}^\dag }{{\bf{C}}_{ - A}}{\bf{F}}{{\bf{C}}_{ - \frac{1}{A}}}{{\bf{F}}^\dag }{\bf{x}}\right\}\label{eq:Bis019_2}.
\end{align}
Accordingly, we make the assumption that (\ref{eq:Bis012}) is used when 
$|A|>|D|$ and (\ref{eq:Bis016}) is used when 
$|A|<|D|$:
\begin{align}\label{eq:Bis020}
&{\cal O}_{{\textmd{DLCT}}}^{(A,0;C,D)}\left\{ {\bf{x}} \right\} \nn\\
&=\left\{
\begin{array}{ll}
\sqrt{-j}\ {\bf{F}} {{\bf{C}}_{\frac{1}{D}}}{{\bf{F}}^\dag }{{\bf{C}}_D}{\bf{F}}{{\bf{C}}_{\frac{{C + 1}}{D}}}{\bf{x}},&\textmd{for }\ |A|>|D|\\
\sqrt{j}\ {{\bf{C}}_{\frac{{C - 1}}{A}}}{{\bf{F}}^\dag }{{\bf{C}}_{ - A}}{\bf{F}}{{\bf{C}}_{ - \frac{1}{A}}}{{\bf{F}}^\dag }{\bf{x}},&\textmd{for }\ |A|<|D|
\end{array}
\right..
\end{align}
(Alternatively, one can choose using another assumption that (\ref{eq:Bis012}) is used when $|A|<|D|$ and (\ref{eq:Bis016}) is used when $|A|>|D|$.)
For example, consider $(A,B;C,D)=(2,0;1,0.5)$ and then $(A,B;C,D)^{-1}=(0.5,0;-1,2)$.
The DLCT with $(2,0;1,0.5)$ uses the form for $|A|>|D|$ in (\ref{eq:Bis020}) because $2>0.5$.
The IDLCT, i.e. DLCT with $(0.5,0;-1,2)$, uses the form for $|A|<|D|$ in (\ref{eq:Bis020}) because $0.5<2$.
The reversibility holds obviously because
\begin{align}\label{eq:Bis024}
{\bf{x}} &={\cal O}_{{\textmd{DLCT}}}^{(0.5,0;-1,2)}\left\{ {{\cal O}_{{\textmd{DLCT}}}^{(2,0;1,0.5)}\left\{ {\bf{x}} \right\}} \right\}\nn\\
&=\sqrt{j}\ {{\bf{C}}_{-4}}{{\bf{F}}^\dag }{{\bf{C}}_{ - 0.5}}{\bf{F}}{{\bf{C}}_{ -2}}{{\bf{F}}^\dag }\left( \sqrt{-j}\ {\bf{F}}{{\bf{C}}_{2}}{{\bf{F}}^\dag }{{\bf{C}}_{0.5}}{\bf{F}}{{\bf{C}}_{4}{\bf{x}}} \right).
\end{align}

\subsection{Special Cases of the DLCT}\label{subsec:Spec}
In this subsection, we discuss some special cases of the proposed DLCT, including DFRFT, discrete  Fresnel transform 
and discrete scaling operation.

\vspace*{6pt}
\subsubsection{Discrete fractional Fourier transform}
In \cite{pei2002eigenfunctions}, it has been shown that the FRFT of fractional angle $\alpha$ is the special case of the LCT of ${\bb M}=(\cos\alpha,\sin\alpha;-\sin\alpha,\cos\alpha)$ with some phase difference:
\begin{align}\label{eq:Spec04}
{\cal O}_{{\textmd{FRFT}}}^{\alpha}\left\{ x(t) \right\}
=e^{j\frac{\alpha}{2}}{\cal O}_{{\textmd{LCT}}}^{(\cos\alpha,\sin\alpha;-\sin\alpha,\cos\alpha)}\left\{ x(t) \right\}.
\end{align}
Discard the case that $\alpha=0$, i.e. ${\bb M}=(1,0;0,1)$, because it is just an identity operation.
Then, the DFRFT can be implemented by the proposed DLCT defined in (\ref{eq:Bnot040}) with the same phase difference $e^{j\frac{\alpha}{2}}$.
That is,
\begin{align}\label{eq:Spec08}
{\cal O}_{{\textmd{DFRFT}}}^{\alpha}\left\{ \bb x\right\}
& \buildrel \Delta \over = e^{j\frac{\alpha}{2}}{\cal O}_{{\textmd{DLCT}}}^{(\cos\alpha,\sin\alpha;-\sin\alpha,\cos\alpha)}\left\{ \bb x \right\}\nn\\
&= e^{j\frac{\alpha}{2}}{{\bf{C}}_{\frac{{\cos\alpha - 1}}{\sin\alpha}}}{{\bf{F}}^\dag }{{\bf{C}}_{ - \sin\alpha}}{\bf{F}}{{\bf{C}}_{\frac{{\cos\alpha - 1}}{\sin\alpha}}}{\bf{x}}\nn\\
&= e^{j\frac{\alpha}{2}}{{\bf{C}}_{-\tan\frac{\alpha}{2}}}{{\bf{F}}^\dag }{{\bf{C}}_{ - \sin\alpha}}{\bf{F}}{{\bf{C}}_{-\tan\frac{\alpha}{2}}}{\bf{x}}.
\end{align}

In \cite{ozaktas1996digital}, the authors proposed two digital computation methods for the FRFT.
The second method is similar to the direct summation method in (\ref{eq:Bnot004}) with $(A,B;C,D)=(\cos\alpha,\sin\alpha;-\sin\alpha,\cos\alpha)$ and $N=(2\Delta_x)^2$.
The first method presents similar notion as that in (\ref{eq:Spec08}), i.e. decomposing the FRFT into CM-CC-CM.
However, the difference between them is the implementation of the CC.
In \cite{ozaktas1996digital}, the CC is digitally implemented by a discrete linear convolution and is recommended using FFTs.
So the inputs of the convolution have to be padded with zeros before passing through FFTs.
The CC in (\ref{eq:Spec08}) is digitally implemented by two FFTs and one discrete CM, like a discrete circular convolution without zero-padding.

\vspace*{6pt}
\subsubsection{Discrete Fresnel transform}
The 1-D  Fresnel transform with wavelength $\lambda$ and propagation distance $z$,
denoted by ${\cal O}_{{\textmd{Fresnel}}}^{\lambda,z}$, is also a special case of the LCT where ${\bb M}=(1,\lambda z;0,1)$ \cite{pei2002eigenfunctions}, i.e. a CC,
\begin{align}\label{eq:Spec12}
{\cal O}_{{\textmd{Fresnel}}}^{\lambda,z}\left\{ x(t) \right\}
=e^{j\frac{\pi z}{\lambda}}{\cal O}_{{\textmd{LCT}}}^{(1,\lambda z;0,1)}\left\{ x(t) \right\}.
\end{align}
Accordingly, we can design the discrete Fresnel transform
${\cal O}_{{\textmd{DFresnel}}}^{\lambda,z}$ as follows using the proposed DLCT defined in (\ref{eq:Bnot040}):
\begin{align}\label{eq:Spec16}
{\cal O}_{{\textmd{DFresnel}}}^{\lambda,z}\left\{\bb x \right\}
& \buildrel \Delta \over = e^{j\frac{\pi z}{\lambda}}{\cal O}_{{\textmd{DLCT}}}^{(1,\lambda z;0,1)}\left\{ \bb x \right\}\nn\\
&= e^{j\frac{\pi z}{\lambda}} {{\bf{F}}^\dag }{{\bf{C}}_{ - \lambda z}}{\bf{F}}\bb x.
\end{align}

\vspace*{6pt}
\subsubsection{Discrete scaling}
When $\bb M=(\sigma,0;\sigma^{-1},0)$, the LCT reduces to the scaling operation with scaling parameter $\sigma$, denoted by ${\cal O}_{{\textmd{Scal}}}^{\sigma}$:
\begin{align}\label{eq:Spec52}
{\cal O}_{{\textmd{Scal}}}^{\sigma}\left\{ x(t) \right\}
=x( t/\sigma)
={\cal O}_{{\textmd{LCT}}}^{(\sigma,0;\sigma^{-1},0)}\!\left\{ x(t) \right\}.
\end{align}
Recall the proposed DLCT for $B=0$ shown in 
(\ref{eq:Bis020}).
The discrete scaling operation ${\cal O}_{{\textmd{DScal}}}^{\sigma}$ is given by
\begin{align}\label{eq:Spec56}
{\cal O}_{{\textmd{DScal}}}^{\sigma}\!\left\{ \bb x \right\}\!
=\! \left\{ {\begin{array}{*{20}{c}}
{\!\sqrt { - j} \;{\bf{F}}{{\bf{C}}_\sigma }{{\bf{F}}^\dag }{{\bf{C}}_{\frac{1}{\sigma }}}{\bf{F}}{{\bf{C}}_\sigma }{\bf{x}},}&{\left| \sigma  \right| > 1}\\
{\!\sqrt j \;{{\bf{C}}_{ - \frac{1}{\sigma }}}{{\bf{F}}^\dag }{{\bf{C}}_{ - \sigma }}{\bf{F}}{{\bf{C}}_{ - \frac{1}{\sigma }}}{{\bf{F}}^\dag }{\bf{x}},}&{\left| \sigma  \right| < 1}
\end{array}} \right..
\end{align}
As mentioned in 
(\ref{eq:Bis018_1})-(\ref{eq:Bis019_2}), the above two kinds of decompositions are used for perfect reversibility.

\vspace*{6pt}
\subsubsection{Other discrete operations}
The FT, inverse FT and CM are also the special cases of the LCT with parameter matrix being $(0,1;-1,0)$, $(0,-1;1,0)$ and $(1,0;\xi,1)$, respectively.
The discrete versions of these operations are simply the DFT ($\bf{F}$), IDFT (${\bf{F}}^\dag$) and discrete CM (${\bf{C}}_\xi$), respectively, without the need of the proposed DLCT.

\section{Relation Between Proposed DLCT and Continuous LCT}\label{sec:Relation}
In this section, we discuss the connections between the proposed DLCT and continuous LCT, including derivation of the DLCT from the continuous LCT, relation between $(A,B;C,D)$ of the DLCT and  $(a,b;c,d)$ of the continuous LCT, and oversampling for the DLCT to approximate the samples of the continuous LCT.
Some related works regarding sampling and oversampling of the LCT include \cite{stern2006sampling,ozaktas2006efficient,li2007new,koc2008digital,healy2009sampling}.
In this paper, we discuss sampling and oversampling from the point of view of CM-CC-CM decomposition.

\subsection{Derivation of Proposed DLCT From Continuous LCT}\label{sec:Rel}
Since the proposed DLCT is based on the CM-CC-CM decomposition,
consider the expression of the continuous LCT based on the same decomposition:
\begin{align}\label{eq:Rel04}
&\textmd{(CM)}\ x_1(t)={e^{j\pi \frac{{a - 1}}{b}{t^2}}}x(t) ,\\
&\textmd{(CC)}\ {X_1}(u) = \int\limits_{ - \infty }^\infty  {{e^{ - j\pi b{f^2}}}\!\left( \int\limits_{ - \infty }^\infty  {{x_1}(t){e^{ - j2\pi ft}}}  dt\right)} {e^{j2\pi fu}}df \label{eq:Rel05},\\
&\textmd{(CM)}\ X(u)={e^{j\pi \frac{{d - 1}}{b}u^2}}X_1(u) \label{eq:Rel06}.
\end{align}
Let ${\widehat x}_1(f)$ and ${\widehat X}_1(f)$ denote the FTs of $x_1(t)$ and ${X_1}(u)$, respectively:
\begin{align}\label{eq:Rel07}
{\widehat x}_1(f)={\cal F}\left\{x_1(t)\right\}\quad\textmd{and}\quad
{\widehat X}_1(f)={\cal F}\left\{X_1(u)\right\},
\end{align}
where ${\cal F}$ denotes the FT operation.
From (\ref{eq:Rel07}) and (\ref{eq:Rel05}), it follows that
\begin{align}\label{eq:Rel08}
{\widehat X}_1(f)= e^{ - j\pi b{f^2}}{\widehat x}_1(f).
\end{align}
Sample $x_1(t)$ and ${X_1}(u)$  with the same sampling period $\Delta_t=\Delta_u=\Delta$.
Replacing $x_1(t)$ and ${X_1}(u)$ in (\ref{eq:Rel07}) by their discrete samples $x_1(n\Delta)$ and $X_1(k\Delta)$ leads to
\begin{align}\label{eq:Rel09_1}
{{\widehat x}_{1,\frac{1}{{{\Delta}}}}}(f)
 \buildrel \Delta \over = \sum\limits_{l =  - \infty }^\infty\! {{{\widehat x}_1} \left( {f - \frac{l}{{{\Delta}}}} \right)}=  {\Delta}\!\!\sum\limits_{n =  - \infty }^\infty\!  {{x_1}(n\Delta){e^{ - j2\pi fn{\Delta}}}} ,
\end{align}
\begin{align}\label{eq:Rel09_2}
{{\widehat X}_{1,\frac{1}{{{\Delta}}}}}(f)
 \buildrel \Delta \over =  \sum\limits_{l =  - \infty }^\infty\!  {{{\widehat X}_1} \left( {f - \frac{l}{{{\Delta}}}} \right)}
 = {\Delta}\!\!\sum\limits_{k =  - \infty }^\infty\!  {{X_1}(k\Delta){e^{ - j2\pi fk{\Delta}}}}.
\end{align}
To further simplify the above computations, we sample ${{\widehat x}_{1,\frac{1}{{{\Delta}}}}}(f)$ and ${{\widehat X}_{1,\frac{1}{{{\Delta}}}}}(f)$ by sampling period
$\Delta_f=\frac{1}{N\Delta}$ where $N$ is some positive integer.
Then, with $f=\frac{m}{N\Delta}$, (\ref{eq:Rel09_1}) becomes
\begin{align}\label{eq:Rel11}
{{\widehat x}_{1,\frac{1}{{{\Delta}}}}}\left( {\frac{m}{{N\Delta }}} \right)
&=  {\Delta}\sum\limits_{n =  - \infty }^\infty   {{x_1}(n\Delta){e^{ - j\frac{2\pi}{N} mn}}}\nn\\
&= {\Delta}\sum\limits_{n =  - N/2 }^{N/2-1} {x_{1,N\Delta}}\left(n\Delta\right){e^{ - j\frac{2\pi}{N} mn}},
\end{align}
where ${x_{1,N\Delta}}(t)$ is the periodic summation of $x_1(t)$ with period $N\Delta$:
\begin{align}\label{eq:Rel11_2}
{x_{1,N\Delta}}(t) \buildrel \Delta \over = \sum\limits_{l =  - \infty }^\infty   {x_1}(t-lN\Delta),
\end{align}
Similarly, sampling (\ref{eq:Rel09_2}) with $f=\frac{m}{N\Delta}$ leads to
\begin{align}\label{eq:Rel12}
{{\widehat X}_{1,\frac{1}{{{\Delta}}}}}\left( {\frac{m}{{N\Delta }}} \right)
&= {\Delta}\sum\limits_{k =  - N/2 }^{N/2-1}   {X_{1,N\Delta}}\left(k\Delta\right){e^{ - j\frac{2\pi}{N} mk}},
\end{align}
where ${X_{1,N\Delta}}(u)$ is the periodic summation of $X_1(u)$:
\begin{align}\label{eq:Rel12_2}
{X_{1,N\Delta}}(u) \buildrel \Delta \over = \sum\limits_{l =  - \infty }^\infty   {X_1}(u-lN\Delta).
\end{align}
The inverse transform of (\ref{eq:Rel12}) is given by
\begin{align}\label{eq:Rel14}
 {X_{1,N\Delta}}\left(k\Delta\right)
= \frac{1}{N\Delta} \sum\limits_{m =  - N/2 }^{N/2-1}{{\widehat X}_{1,\frac{1}{{{\Delta}}}}}\left( {\frac{m}{{N\Delta }}} \right) {e^{ j\frac{2\pi}{N} mk}}.
\end{align}
Finally, a DLCT can be developed by the following five steps
(the combination of the 2nd to 4th steps corresponds to the CC procedure):
\begin{align}\label{eq:Rel15_1}
x_{1,N\Delta}(n\Delta)&={e^{j\frac{\pi}{N} \frac{{a - 1}}{b}N\Delta^2{n^2}}}x(n\Delta) ,\\
{{\widehat x}_{1,\frac{1}{{{\Delta}}}}}\left( {\frac{m}{{N\Delta }}} \right)
&= {\Delta}\sum\limits_{n =  - N/2 }^{N/2-1}   {x_{1,N\Delta}}(n\Delta){e^{ - j\frac{2\pi}{N} mn}}\label{eq:Rel15_2},\\
{{\widehat X}_{1,\frac{1}{{{\Delta}}}}}\left( {\frac{m}{{N\Delta }}} \right)
&={e^{ - j\frac{\pi}{N} b\frac{1}{N\Delta^2}{m^2}}}{{\widehat x}_{1,\frac{1}{{{\Delta}}}}}\left( {\frac{m}{{N\Delta }}} \right)\label{eq:Rel15_3},
\end{align}
\begin{align}
{X_{1,N\Delta}}\left(k\Delta\right)
&= \frac{1}{N\Delta} \sum\limits_{m =  - N/2 }^{N/2-1}{{\widehat X}_{1,\frac{1}{{{\Delta}}}}}\!\left( {\frac{m}{{N\Delta }}} \right) {e^{ j\frac{2\pi}{N} mk}}\label{eq:Rel15_4},\\
X[k] &= {e^{j\frac{\pi}{N} \frac{{d - 1}}{b}N\Delta^2 k^2}}{X_{1,N\Delta}}\left(k\Delta\right)\label{eq:Rel15_5},
\end{align}
where $-\frac{N}{2}\leq n,m,k\leq\frac{N}{2}-1$.
The assumptions in (\ref{eq:Rel15_1}), (\ref{eq:Rel15_3}) and (\ref{eq:Rel15_5}) may be inconsistent with the relations in (\ref{eq:Rel04}), (\ref{eq:Rel08}) and (\ref{eq:Rel06}).
If so, the output $X[k]$ of the DLCT will be different from the sampled output $X(k\Delta)$ of the continuous LCT.
In next subsection, we will discuss the conditions such that the above DLCT can produce an accurate approximation to $X(k\Delta)$.

Compared with (\ref{eq:Bnot028})-(\ref{eq:Bnot036}), the above DLCT  is equivalent to the proposed DLCT if $ \frac{{a - 1}}{b}N\Delta^2= \frac{{A - 1}}{B}$, $b\frac{1}{N\Delta^2}=B$ and $\frac{{d - 1}}{b}N\Delta^2=\frac{{D - 1}}{B}$.
It implies that
the relation between the parameter matrix $(A,B;C,D)$ of the proposed DLCT and the parameter matrix $(a,b;c,d)$ of the continuous LCT is given by
\begin{align}\label{eq:Rel40}
\begin{bmatrix}
A&B\\
C&D
\end{bmatrix}
=\begin{bmatrix}
a&{\frac{b}{{N{\Delta^2}}}}\\
{cN{\Delta^2}}&d
\end{bmatrix},
\end{align}
where $(A,B;C,D)=(a,b;c,d)$ when $\Delta=\sqrt{1/N}$.

\subsection{Oversampling for Approximating the Samples of Continuous LCT}\label{subsec:over}
In (\ref{eq:Bnot048}), (\ref{eq:Bis019_1}) and (\ref{eq:Bis019_2}), it has been proved that the proposed DLCT is always reversible.
However, the output of the proposed DLCT may be different from the sampled output of the continuous LCT because of some aliasing and overlapping problems.
Recall the DLCT in (\ref{eq:Rel15_1})-(\ref{eq:Rel15_5}).
If we want $X[k]\approx X(k\Delta)$, 
(\ref{eq:Rel15_1}), (\ref{eq:Rel15_3}) and (\ref{eq:Rel15_5}) have to be consistent with (\ref{eq:Rel04}), (\ref{eq:Rel08}) and (\ref{eq:Rel06}), respectively.
That is, for $-\frac{N}{2}\leq n,m,k\leq\frac{N}{2}-1$, we need
\begin{align}\label{eq:over02_1}
x_{1,N\Delta}(n\Delta)
&=\sum\limits_{l =  - \infty }^\infty   {x_1}\left(n\Delta-lN\Delta\right)=x_{1}(n\Delta) ,\\
{{\widehat x}_{1,\frac{1}{{{\Delta}}}}}\left( {\frac{m}{{N\Delta }}} \right)
&=\sum\limits_{l =  - \infty }^\infty\!  {{{\widehat x}_1} \left( { \frac{m}{{{N\Delta}}}-\frac{l}{{{\Delta}}}} \right)}\approx
{{\widehat x}_1}\left( {\frac{m}{{N\Delta }}} \right),\label{eq:over02_2}\\
{{\widehat X}_{1,\frac{1}{{{\Delta}}}}}\left( {\frac{m}{{N\Delta }}} \right)
&=\sum\limits_{l =  - \infty }^\infty\!  {{{\widehat X}_1}\! \left( { \frac{m}{{{N\Delta}}}- \frac{l}{{{\Delta}}}} \right)}\!\approx
{{\widehat X}_1}\!\left( {\frac{m}{{N\Delta }}} \right),\label{eq:over02_3}\\
{X_{1,N\Delta}}\left(k\Delta\right)
&= \sum\limits_{l =  - \infty }^\infty   {X_1}\left(k\Delta-lN\Delta\right)={X_{1}}\left(k\Delta\right)\label{eq:over02_4}.
\end{align}
The above four conditions imply that $x_1(t)$ and $X_1(t)$ have to be time-limited to $[-\frac{N\Delta}{2}, \frac{N\Delta}{2}]$ and approximately band-limited to $[-\frac{1}{2\Delta},\frac{1}{2\Delta}]$.
(A time-limited signal cannot be completely band-limbed.)
According to (\ref{eq:Rel04}),
$x_1(t)$ has the same time duration as $x(t)$.
And according to (\ref{eq:Rel08}), $x_1(t)$ and $X_1(t)$ have the same bandwidth.
Therefore, $N$ and $\Delta$ need to satisfy the following  conditions such that the output of the DLCT can approximate the sampled output of the continuous LCT:
\begin{align}\label{eq:over06_1}
x(t)&=0\ \ \textmd{for}\ \ |t|\geq\frac{N\Delta}{2},\\
{\widehat x}_1(f)\buildrel \Delta \over ={\cal F}\{x_1(t)\}&\approx0\ \ \textmd{for}\ \ |f|\geq\frac{1}{2\Delta},\label{eq:over06_2}\\
X_1(u)&=0\ \ \textmd{for}\ \ |u|\geq\frac{N\Delta}{2}.\label{eq:over06_3}
\end{align}
It can be found that the bandwidths of $x(t)$ and $X(u)$ are not taken into account.
However, if one wants to reconstruct $X(u)$ from the DLCT output $X[k]$, it is required that
\begin{align}\label{eq:over10}
{\cal F}\left\{X(u)\right\}\approx0\ \ \textmd{for}\ \ |f|\geq\frac{1}{2\Delta}.
\end{align}
Plus, if one wants to recover $x(t)$ from $x[n]$ after some inverse DLCT work, sampling period $\Delta$ should also satisfy
\begin{align}\label{eq:over14}
{\cal F}\left\{x(t)\right\}\approx0\ \ \textmd{for}\ \ |f|\geq\frac{1}{2\Delta}.
\end{align}

\begin{figure}[t]
\centering
\includegraphics[width=.95\columnwidth,clip=true]{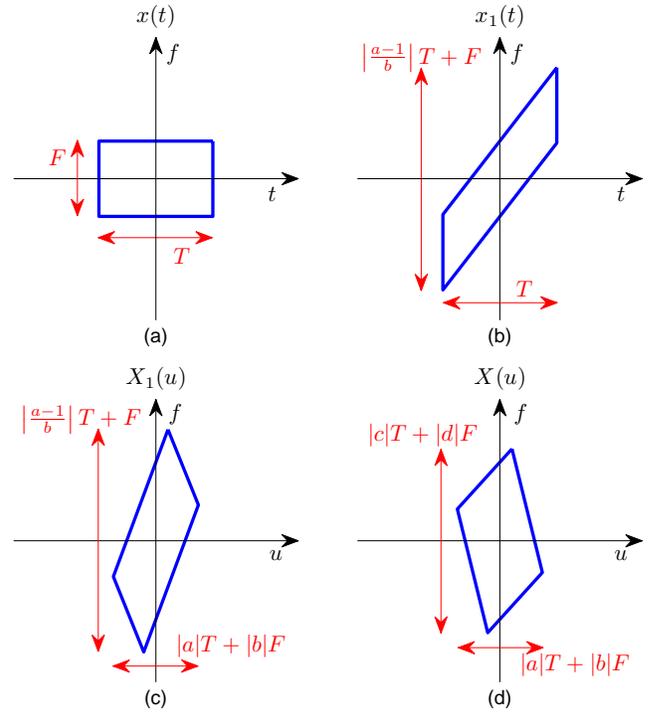}
\vspace*{-4pt}
\caption{
Effect of each step of the LCT with $\bb M=(a,b;c,d)$ in time-frequency plane: (a) the original signal $x(t)$, (b) $x_1(t)$ i.e. $x(t)$ after the CM step, (c) $X_1(u)$ i.e. $x_1(t)$ after the CC step and (d) $X(u)$ i.e. $X_1(u)$ after the CM step.
$T$ and $F$ are the  time duration and bandwidth of $x(t)$, respectively.
}
\label{fig:oversampling_1}
\vspace*{-10pt}
\end{figure}

Consider that the input signal $x(t)$ has time duration $T$ and bandwidth $F$, as shown in Fig.~\ref{fig:oversampling_1}(a).
In \cite{pei2001relations,koc2008digital}, it has been shown that the relation of the LCT to the Wigner distribution function (WDF) is given by
\begin{align}\label{eq:over18}
{W_Y}\left( {u,v} \right) = {W_y}\left( {du - bv, - cu + av} \right),
\end{align}
where ${W_Y}$ and ${W_y}$ are the WDFs of
$Y(u)$ and $y(t)$, respectively, and $Y(u)$ is the LCT of $y(t)$ with parameter matrix $\bb M=(a,b;c,d)$.
The CM with chirp rate $(a-1)/b$, i.e. $\bb M=\left(1,0;(a-1)/b,1\right)$, will lead to shearing in frequency domain.
Therefore, the bandwidth of $x_1(t)$ become $\left|\frac{a-1}{b}\right|T+F$ while the time duration remains the same as $x(t)$, as shown in Fig.~\ref{fig:oversampling_1}(b).
In contrast, the CC, i.e. $\bb M=\left(1,b;0,1\right)$, will lead to shearing in time domain and change the time duration to $|a|T+|b|F$, as shown in Fig.~\ref{fig:oversampling_1}(c).
Fig.~\ref{fig:oversampling_1}(d) shows that the CM changes the bandwidth again to $|c|T+|d|F$.
With these information of the time durations and bandwidths of $x(t)$, $x_1(t)$, $X_1(u)$ and $X(u)$, one can determine the sampling period $\Delta$ and the number of samples $N$.
According to (\ref{eq:over06_2}), $\Delta$ needs to satisfy
\begin{align}\label{eq:over22}
\frac{1}{\Delta}\geq \left|\frac{a-1}{b}\right|T+F.
\end{align}
If the additional conditions (\ref{eq:over10}) and (\ref{eq:over14}) are also considered, we need
\begin{align}\label{eq:over26}
\frac{1}{\Delta}\geq &\max\left\{\left| \frac{a-1}{b}\right|T+F,\ |c|T+|d|F,\ F \right\}\nn\\
&=\max\left\{\left| \frac{a-1}{b}\right|T+F,\ |c|T+|d|F \right\}.
\end{align}
Once $\Delta$ is determined, $N$ can be determined based on (\ref{eq:over06_1}) and (\ref{eq:over06_3}):
\begin{align}\label{eq:over30}
N\geq \frac{1}{\Delta}\max\left\{ T,\ |a|T+|b|F \right\}.
\end{align}
Except when $a=1$, no mater which one of (\ref{eq:over22}) and (\ref{eq:over26}) is adopted, oversampling is required if $x(t)$ is originally sampled based on its bandwidth, i.e. $\Delta=1/F$.

\begin{figure}[t]
\centering
\includegraphics[width= \columnwidth,clip=true]{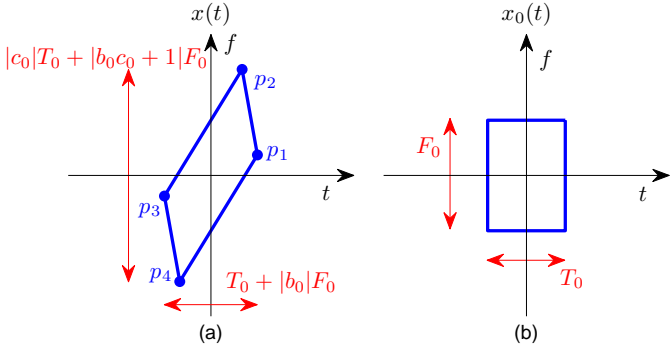}
\vspace*{-4pt}
\caption{
Time-frequency distributions: (a) the original signal $x(t)$; and (b) $x_0(t)$ i.e. the LCT output of $x(t)$ with parameter matrix  $\bb M_0=(1,b_0;c_0,b_0c_0+1)$. The $\bb M_0$ is well-designed such that the time-frequency distribution of $x_0(t)$ is a rectangle with time duration $T_0$ and bandwidth $F_0$.
}
\label{fig:oversampling_2}
\vspace*{-10pt}
\end{figure}

In fact, oversampling would be unnecessary if more time-frequency information of $x(t)$ is provided.
Consider that the time-frequency distribution of $x(t)$ is within some fundamental parallelogram.
For example, assume the time-frequency distribution of $x(t)$ is as shown in Fig.~\ref{fig:oversampling_2}(a).
And assume $x_0(t)$ is the LCT output of $x(t)$ with parameter matrix  $\bb M_0=(1,b_0;c_0,b_0c_0+1)$.
If $\bb M_0$ is well-designed, the time-frequency distribution of $x_0(t)$ will be a rectangle with time duration $T_0$ and bandwidth $F_0$, as shown in Fig.~\ref{fig:oversampling_2}(b).
Given the position information of the four vertices $p_1,p_2,p_3,p_4$ of the  fundamental parallelogram in Fig.~\ref{fig:oversampling_2}(a), say $p_1=(t_1,f_1)$ and $p_2=(t_2,f_2)$, we have
\begin{align}\label{eq:over34}
T_0&=t_1-t_2,&\ \ c_0&=\frac{f_1-f_2}{T_0},\nn\\
F_0&=f_1+f_2-c_0(t_1+t_2),&\ \ b_0&=\frac{t_1+t_2}{F_0}.
\end{align}
Then, according to (\ref{eq:over18}), the time duration $T$ and bandwidth $F$ of $x(t)$ are given by
\begin{align}\label{eq:over36}
T&=T_0+|b_0|F_0,\\
F&=|c_0|T_0+|b_0c_0+1|F_0.\label{eq:over37}
\end{align}
Since $x(t)={\cal O}_{\textmd{LCT}}^{\bb M_0}\{x_0(t)\}$, the relation of $X(u)$ and $x_0(t)$ is
\begin{align}\label{eq:over38}
X(u)= {\cal O}_{\textmd{LCT}}^{\bb M}\{x(t)\}={\cal O}_{\textmd{LCT}}^{\bb M\cdot\bb M_0}\{x_0(t)\}.
\end{align}
Assume $\bb M_1=\bb M\cdot\bb M_0$, i.e.
\begin{align}\label{eq:over40}
\bb M_1=\begin{bmatrix}
a_1\ &b_1\\
c_1\ &d_1
\end{bmatrix} \buildrel \Delta \over =
\begin{bmatrix}
a\ &b\\
c\ &d
\end{bmatrix}
\begin{bmatrix}
1\ &b_0\\
c_0\ &b_0c_0+1
\end{bmatrix}.
\end{align}
In Fig.~\ref{fig:oversampling_1}, replacing $x(t)$ by $x_0(t)$ and $(a,b;c,d)$ by $(a_1,b_1;c_1,d_1)$, the time durations and bandwidths of $x_1(t)$, $X_1(u)$ and $X(u)$ can be expressed in terms of $T_0$ and $F_0$.
Then, (\ref{eq:over22}), (\ref{eq:over26}) and (\ref{eq:over30}) become
\begin{align}\label{eq:over42}
\frac{1}{\Delta}&\geq \left|\frac{a_1-1}{b_1}\right|T_0+F_0,\\
\frac{1}{\Delta}&\geq \max\left\{\left| \frac{a_1-1}{b_1}\right|T_0+F_0,\ |c_1|T_0+|d_1|F_0,\ F \right\},\label{eq:over46}\\
N&\geq \frac{1}{\Delta}\max\left\{ T,\ |a_1|T_0+|b_1|F_0 \right\},\label{eq:over50}
\end{align}
respectively, where $T$ and $F$ are given in (\ref{eq:over36}) and (\ref{eq:over37}).
Note that the constraint (\ref{eq:over46}) instead of (\ref{eq:over42}) is used when one wants to reconstruct $X(u)$ from $X[k]$ and $x(t)$ from $x[n]$.
It is possible that $F$ is equal to or larger than the right-hand sides of (\ref{eq:over42}) and (\ref{eq:over46})
when the input $x(t)$ has larger bandwidth than the intermediate result $x_1(t)$ after the CM step and the final LCT output $X(u)$.
If so, oversampling is unnecessary.

We can summarize the results in (\ref{eq:over26}) and (\ref{eq:over46}) simply as follows.
Let $F$, $F_1$ and $F_{\textmd{out}}$ denote the bandwidths of the input signal $x(t)$, intermediate result $x_1(t)$ after the CM step and the LCT output $X(u)$, respectively.
\begin{itemize}
\item If $F< F_1$ or $F< F_{\textmd{out}}$, then oversampling is required due to LCT bandwidth expansion.
\item If $F\geq F_1$ and $F\geq F_{\textmd{out}}$, then oversampling is not required due to LCT bandwidth compression.
\end{itemize}
If one wants to calculate the samples of the continuous LCT by the proposed DLCT, once the sampling period $\Delta$ and the number of samples $N$ are determined, the parameter matrix of the proposed DLCT can be obtained from (\ref{eq:Rel40}), i.e. $(A,B;C,D)=(a, b/(N\Delta^2);cN\Delta^2,d)$.

Note that oversampling for decreasing $\Delta$ and zero-padding for increasing $N$ are classified as data preprocessors, not parts of the proposed DLCT, because they are totally unnecessary in some applications such as data encryption and decryption.

\begin{table}[t]
\footnotesize
\begin{center}
\setstretch{1.5}
\caption{Complexity of direct summation method, CDDHFs-based DLCT \cite{pei2011discrete} and proposed DLCT}\label{tab:Comp}
\begin{tabular}{|c|c|c|}
\hline
 &Matrix form & Complexity\textsuperscript{$1$} \\
\hline\hline
Direct sum & $\bb K\bb x$ & $4N^2$\\
\hline
CDDHFs-based\textsuperscript{$2$} & ${{\bf{C}}_\xi }{{\bf{E}}_\sigma }{{\bf{V}}_\alpha }{\bf{E}}_1^T{\bf{x}}$ & $4N^2+8N$\\
\hline
Proposed ($\!B\!\!\neq\!\!0$)& ${{\bf{C}}_{\frac{{D - 1}}{B}}}{{\bf{F}}^\dag }{{\bf{C}}_{ - B}}{\bf{F}}{{\bf{C}}_{\frac{{A - 1}}{B}}}{\bf{x}}$ & $\!\!12N+4N\log_2N\!\!$\\
\hline
\multirow{2}{*}{Proposed ($\!B\!\!=\!\!0$)}
& $\sqrt{-j}{\bf{F}} {{\bf{C}}_{\frac{1}{D}}}{{\bf{F}}^\dag }{{\bf{C}}_D}{\bf{F}}{{\bf{C}}_{\frac{{C + 1}}{D}}}{\bf{x}}$
& \multirow{2}{*}{$\!\!12N+6N\log_2N\!\!$}\\
& $\!\!\sqrt{j}{{\bf{C}}_{\frac{{C - 1}}{A}}}{{\bf{F}}^\dag }{{\bf{C}}_{ - A}}{\bf{F}}{{\bf{C}}_{ - \frac{1}{A}}}{{\bf{F}}^\dag }{\bf{x}}\!\!$
&  \\
\hline
\multicolumn{3}{l}{
\footnotesize{1. List the number of real multiplications.}}\\
\multicolumn{3}{l}{\footnotesize{2. The complexity of matrix eigendecomposition for ${\bf{E}}_\sigma$ and ${\bf{E}}_1$ is not}}\\
\multicolumn{3}{l}{\footnotesize{\quad  included.}}
\end{tabular}
\end{center}
\vspace*{-18pt}
\end{table}

\section{Comparisons Between Proposed DLCT and CDDHFs-based DLCT \cite{pei2011discrete}}\label{sec:Comp}
In this section, the proposed DLCT will be compared with the previous work, 
CDDHFs-based DLCT in \cite{pei2011discrete}, which is also irrelevant to the sampling periods and without oversampling operation.
In the following, the comparisons in computational complexity, accuracy of the approximation to sampled continuous LCT, additivity property and reversibility property are presented.

\subsection{Computational Complexity}\label{subsec:Complex}
Recall the matrix form of the DLCT based on direct summation (\ref{eq:Bnot002}), i.e. $\bb X= \bb K\bb x$.
It involves $N^2$ complex multiplications, and thus the computational complexity is $4N^2$ real multiplications.
Next, consider the CDDHFs-based DLCT \cite{pei2011discrete} with matrix form shown in (\ref{eq:Lai16}), i.e. ${\bf{X}} = {{\bf{C}}_\xi }{{\bf{E}}_\sigma }{{\bf{V}}_\alpha }{\bf{E}}_1^T{\bf{x}}$.
Each of the real matrices $\bb E_1$ and $\bb E_\sigma$ yields $2N^2$ real multiplications, while each of the complex diagonal matrices ${{\bf{C}}_\xi }$ and ${{\bf{V}}_\alpha }$ leads to $N$ complex multiplications.
Therefore, the computation totally contains $4N^2+8N$ real multiplications.
Regarding the proposed DLCT, firstly consider the $B\neq0$ case with matrix form given in (\ref{eq:Bnot040}): ${\bf{X}} ={{\bf{C}}_{\frac{{D - 1}}{B}}}{{\bf{F}}^\dag }{{\bf{C}}_{ - B}}{\bf{F}}{{\bf{C}}_{\frac{{A - 1}}{B}}}{\bf{x}}$.
There are three discrete CMs which totally require $3N$ complex multiplications.
The DFT ($\bb F$) and IDFT ($\bb F^\dag$) can be computed by the FFT along with $(N/2)\log_2N$ complex multiplications.
Accordingly, the complexity of the proposed DLCT for $B\neq0$ is $12N+4N\log_2N$ real multiplications.
The proposed DLCT for $B=0$ given in (\ref{eq:Bis012}) and (\ref{eq:Bis016}) contains one more DFT (or IDFT) than the $B\neq0$ case.
It follows that the complexity becomes $12N+6N\log_2N$ real multiplications.
At last, we summarize the comparison of the computational complexity in TABLE~\ref{tab:Comp}.

\subsection{Accuracy of Approximation to Sampled LCT}\label{subsec:Acc}
In this subsection, we will examine the 
errors between the DLCTs and the sampled output of the continuous LCT.
In \cite{pei2011discrete}, 
it has been mentioned that the 
continuous LCT of the Gaussian function $g_s(t)=e^{-2\pi st^2}$ is given by
\begin{align}\label{eq:Acc04}
G^{\bb M}_s(u)={\cal O}_{{\textmd{LCT}}}^{(a,b;c,d)}\left\{ g_s(t) \right\}
=\sqrt {\frac{1}{{a + j2bs}}} \,{e^{\frac{{ad - 1 + j2sbd}}{{2\pi s{b^2} - j\pi ab}}{\pi ^2}{u^2}}},
\end{align}
where $b\neq0$.
We adopt the sampling periods $\Delta_t=\Delta_u=\Delta=\sqrt{1/N}$ which are suitable for both 
the proposed DLCT and the CDDHFs-based DLCT \cite{pei2011discrete}.
Then, the sampled input and sampled output of (\ref{eq:Acc04}) are given by
\begin{align}\label{eq:Comp02}
g_s[n]=g_s\left(\frac{n}{\sqrt{N}}\right), \ \   G^{\bb M}_s[k]=G^{\bb M}_s\left(\frac{k}{\sqrt{N}}\right),
\end{align}
respectively.
The accuracy is measured by the normalized mean-square error (NMSE) defined as
\begin{align}\label{eq:Comp04}
{\textmd{NMSE}} = \frac{{\sum\limits_k^{} {{{\left|{{G^{\bb M}_s}[k] - O_{\textmd{DLCT}}^{\bb M}\left\{ {{g_s}[n]} \right\}} \right|}^2}} }}{{\sum\limits_k^{} {{{\left| {{G^{\bb M}_s}[k]} \right|}^2}} }}.
\end{align}
Note that $(A,B;C,D)$ in the proposed DLCT is equal to $(a,b;c,d)$
in the CDDHFs-based DLCT because $\Delta=\sqrt{1/N}$ 
and (\ref{eq:Rel40}).

\begin{figure}[t]
\centering
\includegraphics[width=\linewidth,clip=true]{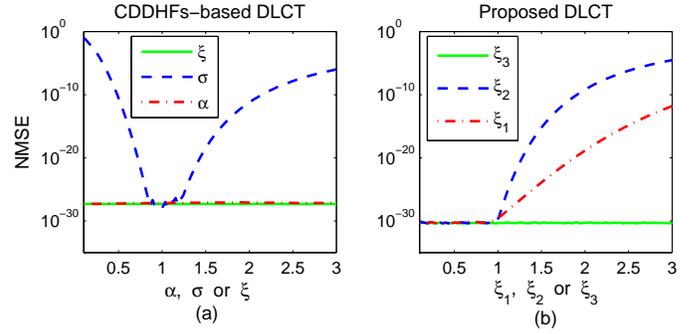}
\vspace*{-18pt}
\caption{
Normalized mean-square error (NMSE) of the approximation to the sampled LCT versus the parameters of (a) CDDHFs-based DLCT and (b) proposed DLCT.
The input is sampled Gaussian function $g_1[n]=e^{-2\pi n^2/N}$.
When one of the parameters varies from $0.1$ to $3$, the other two are fixed to $0.1$ ($\sigma$ to $1.1$).
(a) shows that the scaling parameter $\sigma$ dominates the accuracy of CDDHFs-based DLCT, while (b) shows that the accuracy of the proposed method depends on the first and second chirp rates, i.e. $\xi_1$ and $\xi_2$.
}
\label{fig:ApproxCon_3pars}
\vspace*{-0pt}
\end{figure}
\begin{figure}[t]
\centering
\includegraphics[width=.95\linewidth,clip=true]{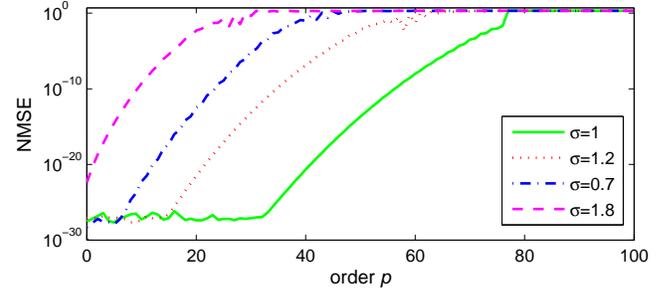}
\vspace*{-6pt}
\caption{
Normalized mean-square error (NMSE) between the CDDHFs (i.e.$\Phi_{p;\sigma}^{\textmd{CDDHF}}$) and the samples of scaled HFs for $N=101$ and scaling parameter $\sigma=0.7,1,1.2,1.8$.
There are $101$ CDDHFs, i.e. $\Phi_{p;\sigma}^{\textmd{CDDHF}}$ of order $p=0,1,\ldots,100$, which form an orthonormal set.
As $\sigma$ is farther from $1$, the error is larger.
}
\label{fig:CDDHFsNMSE}
\vspace*{-10pt}
\end{figure}

The CDDHFs-based DLCT can also be controlled by three independent parameters: DFRFT angle $\alpha$, scaling parameter $\sigma$ and chirp rate $\xi$, as shown in TABLE~\ref{tab:Comp}.
To examine the influence of $\alpha$ on the accuracy, let $\alpha$ vary from $0.1$ to $3$ while $\sigma$ and $\xi$ are fixed to $1.1$ and $0.1$, respectively.
(If $\sigma$ and $\xi$ are fixed to $1$ and $0$, respectively, $b$ will become zero and not suitable for (\ref{eq:Acc04}).)
Similarly, $\alpha$ and $\xi$ are both fixed to $0.1$ when $\sigma$ varies from $0.1$ to $3$.
And $\alpha$ and $\sigma$ are fixed to $0.1$ and $1.1$, respectively, when $\xi$ varies from $0.1$ to $3$.
The NMSEs versus $\alpha$, $\sigma$ and $\xi$ are depicted in Fig.~\ref{fig:ApproxCon_3pars}(a).
Here, $g_s[n]$ with $s=1$ and $N=101$ is used as the input.
We can find out that the accuracy mainly depends on
$\sigma$.
It is because the approximation errors of CDDHFs significantly increase as $\sigma$ is farther from $1$, as shown in Fig.~\ref{fig:CDDHFsNMSE}.
In contrast, the DFRFT angle $\alpha$ has very little influence on the accuracy.
It is because
the DFRFT uses only the undilated DHFs, 
i.e. CDDHFs with $\sigma=1$, which have good approximation.
The chirp rate $\xi$ in the last step won't introduce any error.
The reason
will be discussed later together with the proposed DLCT.

Consider the proposed DLCT with $B\neq0$ because (\ref{eq:Acc04}) is valid only when $b\neq0$
($b=B$ because $\Delta=\sqrt{1/N}$).
As shown in TABLE~\ref{tab:Comp}, the proposed DLCT is actually controlled by three independent chirp rates, i.e. $\xi_1=\frac{{A - 1}}{B}$, $\xi_2=- B$ and $\xi_3=\frac{{D - 1}}{B}$.
Again, we examine the influence of each chirp rate on the accuracy by fixing the other two to $0.1$.
The NMSEs versus $\xi_1$, $\xi_2$ ans $\xi_3$ from $0.1$ to $3$ are shown in Fig.~\ref{fig:ApproxCon_3pars}(b).
As mentioned in Sec.~\ref{subsec:over}, CMs will produce shearing in frequency domain while CC will produce shearing in time domain.
Therefore, larger values of $\xi_1$ and  $\xi_2$ will lead to larger bandwidths of $x_1(t)$ and longer time duration of $X_1(u)$, cause larger errors in the two necessary conditions (\ref{eq:over06_2}) and (\ref{eq:over06_3}) followed by lower accurate DLCT output $X[k]$.
Although the last CM with chirp rate $\xi_3$ will also yield frequency domain shearing, it won't affect the accuracy of $X[k]$ but the accuracy of recovering $X(u)$ from $X[k]$.

\begin{figure}[t]
\centering
\includegraphics[width=.95\linewidth,clip=true]{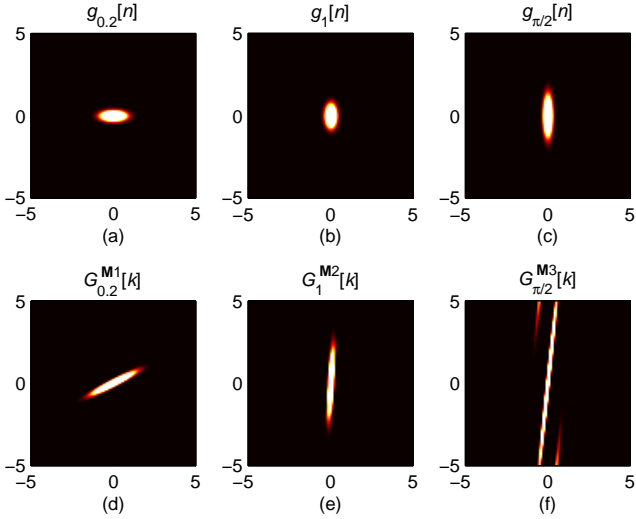}
\vspace*{-6pt}
\caption{
The time-frequency distributions (TFDs) of the
sampled inputs (sampled Gaussian functions) $g_s[n]$ and the sampled LCT outputs $G^{\bb M}_{s}[k]$.
In (a) and (d), $s=0.2$ and $\bb M_1=(0,4;-1/4,2)$ are used.
In (b) and (e), $s=1$ and $\bb M_2=(0.44,-0.08;4.8,1.4)$ are used.
In (c) and (f), $s=\pi/2$ and
$\bb M_3=\left(\frac{0.8}{\sqrt{2}},\frac{0.8}{\sqrt{2}};\frac{5.15}{\sqrt{2}},\frac{7.65}{\sqrt{2}}\right)$ are used.
}
\label{fig:ApproxCon_3cases}
\end{figure}

Recall (\ref{eq:over06_1})-(\ref{eq:over14}) for the proposed DLCT.
The output $X[k]$ without aliasing effect doesn't implies high accuracy of $X[k]$, and vice versa.
This statement is also true for the CDDHFs-based DLCT.
Three examples are presented.
The time-frequency distributions (TFDs) of the sampled Gaussian functions with $s=0.2$, $1$ and $\pi/2$ (i.e. $g_{0.2}[n]$, $g_1[n]$ and $g_{\pi/2}[n]$) are shown in Fig.~\ref{fig:ApproxCon_3cases}(a)-(c).
The TFDs of the sampled LCT outputs with some parameter matrices $\bb M_1$, $\bb M_2$ and $\bb M_3$ (i.e. $G^{\bb M_1}_{0.2}[k]$, $G^{\bb M_2}_1[k]$ and $G^{\bb M_3}_{\pi/2}[k]$) are shown in Fig.~\ref{fig:ApproxCon_3cases}(d)-(f).
There's no aliasing effect in $G^{\bb M_1}_{0.2}[k]$ and $G^{\bb M_2}_1[k]$.
However, the NMSE of $G^{\bb M_1}_{0.2}[k]$ calculated by the CDDHFs-based DLCT is $1.9\times10^{-4}$.
The NMSE of $G^{\bb M_2}_1[k]$ calculated by the proposed DLCT is up to $8.6\times10^{-4}$.
In contrast, $G^{\bb M_3}_{\pi/2}[k]$ has serious aliasing effect, but both the CDDHFs-based and the proposed DLCTs have high accuracy -- NMSEs below $10^{-15}$.

\begin{figure}[t]
\centering
\includegraphics[width=\linewidth,clip=true]{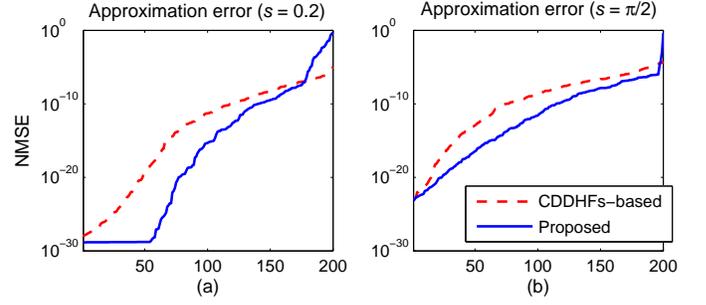}
\vspace*{-18pt}
\caption{
Normalized mean-square errors (NMSEs) of the approximation to the sampled LCT 
for 200 different $\bb M$'s.
The NMSEs are sorted in ascending order.
The input signals are sampled Gaussian functions with (a) $s=0.2$ i.e. $g_{0.2}[n]$ and (b) $s=\pi/2$ i.e. $g_{\pi/2}[n]$.
The parameters in each $\bb M$ are uniformly distributed random numbers on the interval $(-2,2)$.
}
\label{fig:ApproxCon_run200}
\vspace*{-10pt}
\end{figure}

At last, the proposed DLCT is compared with the CDDHFs-based DLCT by 200 simulation runs.
In each run, the  parameters in $\bb M$ are uniformly distributed random numbers on the interval $(-2,2)$.
And we sort the 200 data of NMSEs in ascending order.
The results using $g_{0.2}[n]$ and $g_{\pi/2}[n]$ as the inputs are depicted in Fig.~\ref{fig:ApproxCon_run200}(a) and (b), respectively.
Generally, the proposed method has somewhat higher accuracy than the CDDHFs-based method.

\begin{figure*}[t]
\centering
\includegraphics[width=.9\linewidth,clip=true]{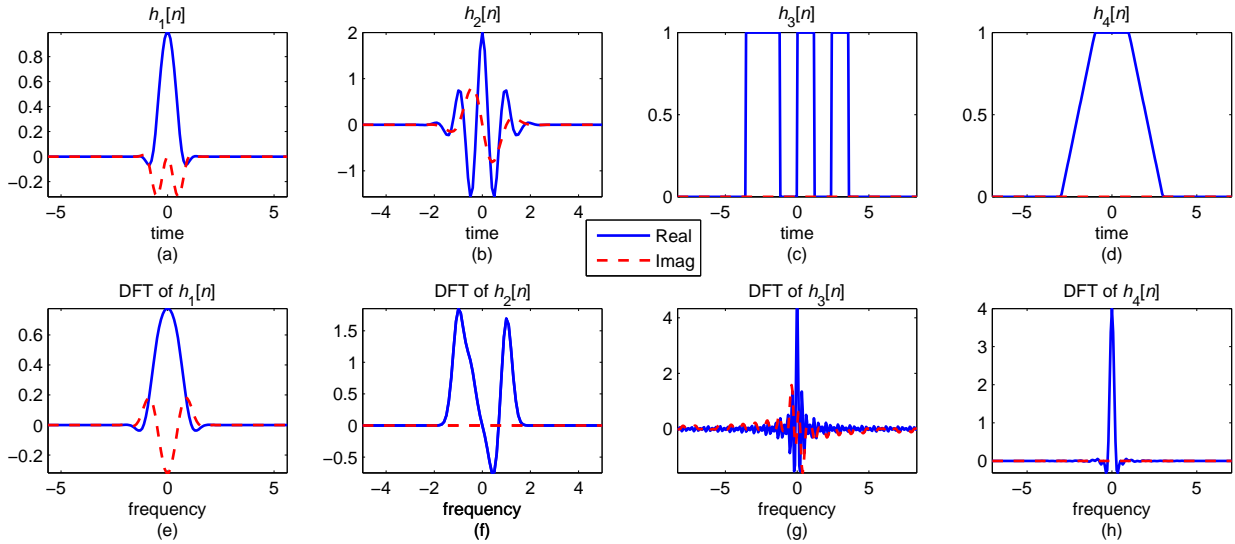}
\vspace*{-6pt}
\caption{
The time domain and frequency domain plots of four kinds of discrete signals $h_1[n]$, $h_2[n]$, $h_3[n]$ and $h_4[n]$  with sampling periods $\sqrt{1/N}$.
Both $h_1[n]$ and $h_2[n]$ are approximately time-limited and band-limited, and $h_1[n]$ has energy more concentrated around the origin then $h_2[n]$. Both $h_3[n]$ and $h_4[n]$ are time-limited but not band-limited, and $h_3[n]$ has a larger bandwidth than $h_4[n]$.
}
\label{fig:Add_input}
\vspace*{-8pt}
\end{figure*}

\subsection{Additivity Property}\label{subsec:Add}
The additivity property for DLCTs is defined as
\begin{align}\label{eq:Add02}
{\cal O}_{\textmd{DLCT}}^{{{\bb M}_1}}{\cal O}_{\textmd{DLCT}}^{{{\bb M}_2}}
= {\cal O}_{\textmd{DLCT}}^{{{\bb M}_1} \times {{\bb M}_2}}.
\end{align}
In the following, we examine the NMSE of the additivity property for the CDDHFs-based and proposed methods:
\begin{align}\label{eq:Add04}
{\textmd{NMSE}} = \frac{{\sum\limits_k^{} {{{\left| {\cal O}_{\textmd{DLCT}}^{{{\bb M}_1} \times {{\bb M}_2}}\left\{ {{h_i}[n]} \right\} - {\cal O}_{\textmd{DLCT}}^{{{\bb M}_1}}{\cal O}_{\textmd{DLCT}}^{{{\bb M}_2}}\left\{ {{h_i}[n]} \right\} \right|}^2}} }}{{\sum\limits_k^{} {{{\left| {\cal O}_{\textmd{DLCT}}^{{{\bb M}_1} \times {{\bb M}_2}}\left\{ {{h_i}[n]} \right\}\right|}^2}} }}.
\end{align}
Four kinds of discrete signals $h_1[n]$, $h_2[n]$, $h_3[n]$ and $h_4[n]$ with sampling periods $\sqrt{1/N}$ are shown in Fig.~\ref{fig:Add_input}(a)-(d) and described as follows:
\begin{itemize}
\item $h_1[n]=e^{-\frac{\pi}{N}n^2-j\frac{\pi}{N}n^2}$ with $N=128$;
\item $h_2[n]=\left[ {2\cos \left( {2\pi \frac{n}{{\sqrt N }}} \right) + j\sin \left( {\pi \left( {\frac{n}{{\sqrt N }} - 1} \right)} \right)} \right]{e^{ - \frac{1}{N}{n^2}}}$ with $N=101$;
\item $h_3[n]$ is a binary sequence with $N=280$;
\item $h_4[n]$ is a trapezoidal-shaped function with $N=201$.
\end{itemize}
The DFTs of these four signals are shown in Fig.~\ref{fig:Add_input}(e)-(h), respectively.

\begin{figure}[t]
\centering
\includegraphics[width=\columnwidth,clip=true]{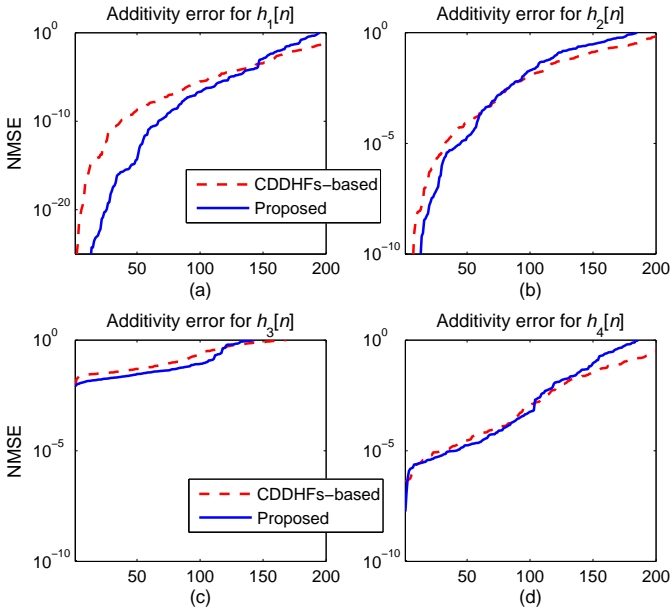}
\vspace*{-18pt}
\caption{
Normalized mean-square errors (NMSEs) of the additivity property 
for 200 different sets of $\bb M_1$ and $\bb M_2$.
The NMSEs are sorted in ascending order.
The input signals are (a) $h_1[n]$, (b) $h_2[n]$, (c) $h_3[n]$ and (d) $h_4[n]$ depicted in Fig.~\ref{fig:Add_input}.
The parameters in $\bb M_1$ and $\bb M_2$ are uniformly distributed random numbers on the interval $(-2,2)$.
}
\label{fig:Add_run200}
\vspace*{-10pt}
\end{figure}

\begin{figure}[t]
\centering
\includegraphics[width=.9\columnwidth,clip=true]{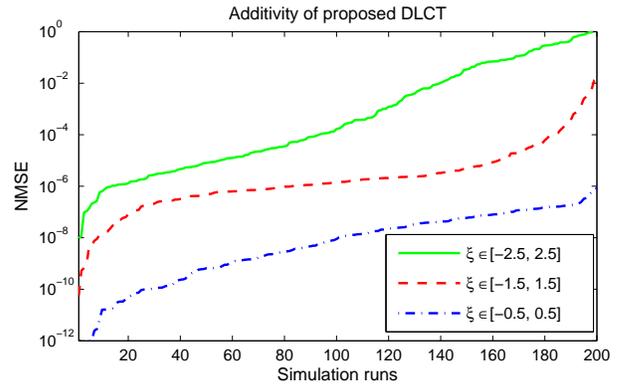}
\vspace*{-6pt}
\caption{
Normalized mean-square errors (NMSEs) of the additivity property of the proposed DLCT with all the chirp rates in ${\cal O}_{\textmd{DLCT}}^{{\bb M}_1}$, ${\cal O}_{\textmd{DLCT}}^{{\bb M}_2}$ and ${\cal O}_{\textmd{DLCT}}^{{{\bb M}_1} \times {{\bb M}_2}}$ are uniformly distributed random numbers within $[-2.5, 2.5]$, $[-1.5, 1.5]$ or $[-0.5, 0.5]$.
The NMSEs are obtained from 200 simulation runs and sorted in ascending order.
}
\label{fig:Add_xi}
\vspace*{-10pt}
\end{figure}

Again, let the parameters in ${\bb M}_1$ and ${\bb M}_2$ be uniformly distributed random numbers on the interval $(-2,2)$, and obtain the additivity NMSE in (\ref{eq:Add04}) for 200 simulation runs.
The NMSEs sorted in ascending order using $h_1[n]$, $h_2[n]$, $h_3[n]$ and $h_4[n]$ as the input are plotted in Fig.~\ref{fig:Add_run200}(a)-(d), respectively.
These four examples reveal that the proposed DLCT has performance similar to the CDDHFs-based DLCT in the additivity property.
Besides, if $N$ is large enough so that the energy is well concentrated around the origin of time-frequency  plane, such as $h_1[n]$, ``approximate'' additivity can be achieved.
When $N$ is not large enough, one way to reduce the NMSE of additivity is to limit the values of $\bb M_1$ and $\bb M_2$.
Choosing all the chirp rates sufficiently small, approximate additivity will be achieved.
For example, use $h_4[n]$ as the input.
With all the chirp rates being uniformly distributed random numbers within $[-2.5, 2.5]$, $[-1.5, 1.5]$ or $[-0.5, 0.5]$, the NMSEs of the additivity
property from 200 simulation runs are plotted in Fig.~\ref{fig:Add_xi}.
It can be found that the NMSE of $\xi\in[-0.5, 0.5]$ is about $10^5\sim10^6$ times smaller than the NMSE of $\xi\in[-2.5, 2.5]$.


\subsection{Reversibility Property}\label{subsec:Rev}
\begin{figure}[t]
\centering
\includegraphics[width=\columnwidth,clip=true]{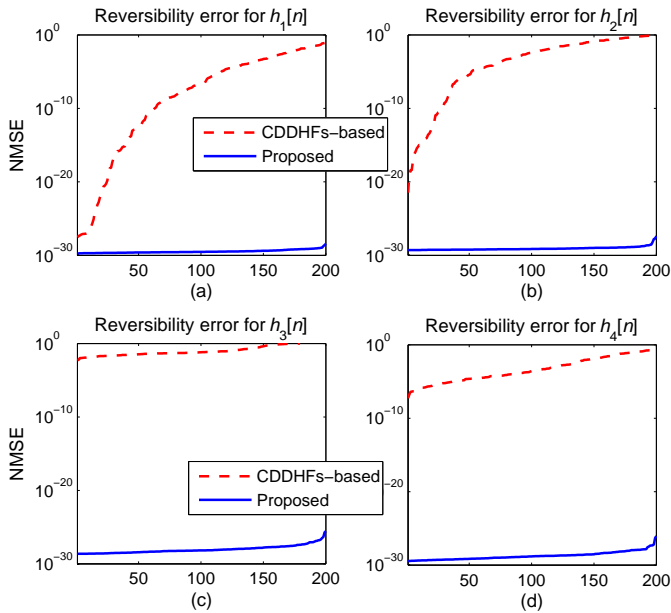}
\vspace*{-18pt}
\caption{
Normalized mean-square errors (NMSEs) of the reversibility property 
for 200 different $\bb M$'s.
The NMSEs are sorted in ascending order.
The input signals are (a) $h_1[n]$, (b) $h_2[n]$, (c) $h_3[n]$ and (d) $h_4[n]$ depicted in Fig.~\ref{fig:Add_input}.
The parameters in each $\bb M$ are uniformly distributed random numbers on the interval $(-2,2)$.
}
\label{fig:Rev_run200}
\vspace*{-2pt}
\end{figure}

Next, we examine the NMSE of the reversibility property:
\begin{align}\label{eq:Rev04}
{\textmd{NMSE}} = \frac{\sum\limits_n^{} {{{\left| h_i[n] - {\cal O}_{\textmd{DLCT}}^{{{\bb M}^{-1}}}{\cal O}_{\textmd{DLCT}}^{{{\bb M}}}\left\{ {{h_i}[n]} \right\} \right|}^2}} }{\sum\limits_n^{} {{{\left| h_i[n]\right|}^2}} }.
\end{align}
Again, 
let the parameters in ${\bb M}$  be uniformly distributed random numbers on the interval $(-2,2)$.
The NMSEs resulting from 200 simulation runs are sorted in ascending order and displayed in Fig.~\ref{fig:Rev_run200}.
The CDDHFs-based method doesn't satisfy the reversibility property perfectly.
Although the proposed DLCT doesn't has perfect additivity, it satisfies the reversibility property perfectly. 
In Fig.~\ref{fig:Rev_run200}, 
all the NMSEs of the proposed DLCT are below $10^{-25}$ and numerically verify the proofs.
With the reversibility property, it is unnecessary to develop the inverse DLCT
additionally because it can be realized by the forward DLCT with ${\bb M}^{-1}$.

At the end, comparisons between the CDDHFs-based DLCT and  the proposed DLCT  are summarized in TABLE~\ref{tab:All}.

\section{Conclusion}\label{sec:Con}
In this paper, we develop a discrete LCT (DLCT) which is irrelevant to the sampling periods and doesn't involve oversampling operation.
The proposed DLCT is based on the well-known CM-CC-CM decomposition, which decomposes the LCT to two chirp multiplications (CMs) and one chirp convolution (CC).
One advantage of this decomposition over many other decompositions is no scaling operation involved because
scaling operation will change the sampling period or introduce interpolation error.
The CM-CC-CM decomposition is invalid for $B=0$.
Accordingly, we modify the decomposition and the proposed DLCT fit for the $B=0$ case.
We also investigate special cases of the proposed DLCT.
The proposed DLCT can be implemented by three discrete CMs and two FFTs (three for $B=0$), which yield lower computational complexity than the previous works, DLCT calculated by direct summation and DLCT based on center discrete dilated Hermite functions (CDDHFs) \cite{pei2011discrete}.
The relation between the proposed DLCT and the continuous LCT is also derived to approximate the samples of the continuous LCT.
Compared with the CDDHFs-based method, the proposed method has somewhat higher 
approximation accuracy.
Besides, simulation results show that approximate additivity property can be achieved with error as small as the CDDHFs-based method.
Most importantly, the proposed method has perfect reversibility, which is proved mathematically and by numerical examples.
With the reversibility property, the inverse transform of the proposed DLCT can be realized by the forward DLCT.

\begin{table}[t]
\footnotesize
\begin{center}
\setstretch{1.5}
\caption{Comparisons between the CDDHFs-based DLCT \cite{pei2011discrete} and proposed DLCT}\label{tab:All}
\begin{tabular}{|c|c|c|}
\hline
 &CDDHFs-based & Proposed \\
\hline\hline
Sampling periods & $\Delta_x=\Delta_u=\sqrt{1/N}$ & $\Delta_x=\Delta_u$\\
\hline
Complexity & $O(N^2)$ & $O(Nlog_2N)$\\
\hline
Accuracy & Worse & Better\\
\hline
Additivity & Approximate & Approximate \\
\hline
Reversibility & Approximate & Perfect\\
\hline
\end{tabular}
\end{center}
\vspace*{-18pt}
\end{table}


\bibliographystyle{IEEEtran}

\begin{IEEEbiography}[{\includegraphics[width=1in,height=1.25in,clip,keepaspectratio]{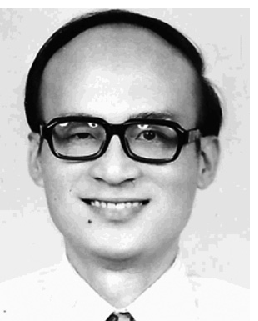}}]
{Soo-Chang Pei} (SM'89-F'00-LF'15) was born in Soo-Auo, Taiwan, China on February 20, 1949. He received the B. S. degree from National Taiwan University in 1970 and the M. S. and Ph. D. degree from the University of California, Santa Barbara in 1972 and 1975 respectively, all in electrical engineering.

He was an engineering officer in the Chinese Navy Shipyard from 1970 to 1971. From 1971 to 1975, he was a research assistant at the University of California, Santa Barbara. He was the Professor and Chairman in the EE department of Tatung Institute of Technology and National Taiwan University, from 1981 to 1983 and 1995 to 1998, respectively. Presently, he is the Professor of EE department at National Taiwan University. His research interests include digital signal processing, image processing, optical information processing, and laser holography. Dr. Pei received National Sun Yet- Sen Academic Achievement Award in Engineering in 1984, the Distinguished Research Award from the National Science Council from 1990-1998, outstanding Electrical Engineering Professor Award from the Chinese Institute of Electrical Engineering in 1998, and the Academic Achievement Award in Engineering from the Ministry of Education in 1998, the IEEE Fellow in 2000 for contributions to the development of digital eigenfilter design, color image coding and signal compression, and to electrical engineering education in Taiwan, the Pan Wen-Yuan Distinguished Research Award in 2002, and the National Chair Professor Award from Ministry of Education in 2002 and 2008. The IEEE Life Fellow in 2015 for recognition of the years of royal membership and support of the activities of IEEE. He has been President of the Chinese Image Processing and Pattern Recognition Society in Taiwan from 1996-1998.

Dr. Pei is IEEE Life Fellow and a member of Eta Keppa Nu and the Optical Society of America.
\end{IEEEbiography}

\begin{IEEEbiography}[{\includegraphics[width=1in,height=1.25in,clip,keepaspectratio]{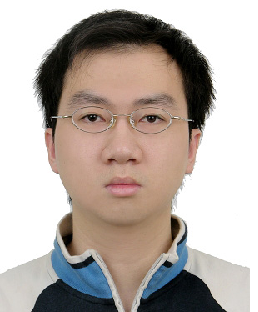}}]
{Shih-Gu Huang} was born in Taiwan in 1984. He received the B.S. degree in electrical engineering and the M.S. degree in communications engineering from National Tsing Hua University, Hsinchu, Taiwan, in 2007 and 2009, respectively. He is currently working toward the Ph.D. degree in the Graduate Institute of Communication Engineering, National Taiwan University, Taipei, Taiwan. His research interests include digital signal processing, time-frequency analysis, fractional Fourier transform, and linear canonical transform.
\end{IEEEbiography}

\vfill

\end{document}